\documentclass[preprint,a4paper,preprintnumbers,amsmath,amssymb,floatfix]{revtex4} 
\usepackage{geometry}                		
\usepackage{graphicx}				
\usepackage[titletoc,toc]{appendix}								
\usepackage{amssymb}
\usepackage{indentfirst}
\usepackage{bm}
\usepackage{amsmath,amsfonts}
\usepackage{float}
\usepackage{subfigure}
\usepackage{xcolor}
\usepackage{array}
\newcolumntype{?}{!{\vrule width 3pt}}
\newcommand{\br}{\bm{r}}
\newcommand{\bn}{\bm{n}}

\newcommand{\mF}{\mathcal{F}}
\newcommand{\bom}{\bm{\omega}}
\newcommand{\be}{\hat{\bm{e}}}

\begin{document}
	
	\title{Osmotic stress and pore nucleation in charged biological nanoshells and capsids}
	\author{Thiago Colla}
	\email{colla@ufop.edu.br}
	\affiliation{Instituto de F\'isica, Universidade Federal de Ouro Preto, CEP 35400-000, Ouro Preto, MG, Brazil}
	\author{Amin Bakhshandeh}
	\email{amin.bakhshandeh@ufrgs.br}
	\affiliation{Programa de P\'os-Gradua\c{c}\~ao em F\'isica, Instituto de F\'isica e Matem\'atica, Universidade Federal de Pelotas, Caixa Postal 354, CEP 96010-900 Pelotas, RS, Brazil}
	\author{Yan Levin} 
	\email{levin@if.ufrgs.br}
	\affiliation{Instituto de F\'isica, Universidade Federal do Rio Grande do Sul,
		Caixa Postal 15051, CEP 91501-970, Porto Alegre, RS, Brazil}

	\begin{abstract}
		A model system is proposed to investigate the chemical equilibrium and mechanical stability of biological spherical-like nanoshells in contact with an aqueous solution with added dissociated electrolyte of a given concentration. The ionic chemical equilibrium across the permeable shell is investigated in the framework of an accurate Density Functional Theory (DFT) that incorporates electrostatic and hardcore correlations beyond the traditional mean-field ({\it e. g.}, Poisson-Boltzmann) limit. The accuracy of the theory is tested by a direct comparison with Monte Carlo (MC) simulations. A simple analytical expression is then deduced which clearly highlights the entropic, electrostatic, and self-energy contributions to the osmotic stress over the shell in terms of the calculated ionic profiles. By invoking a continuum mean-field elastic approach to account for the shell surface stress upon osmotic stretching, the mechanical equilibrium properties of the shell under a wide variety of ionic strengths and surface charges are investigated. The model is further coupled to  a continuum mechanical approach similar in structure to a Classical Nucleation Theory (CNT) to address the question of mechanical stability of the shells against a pore nucleation.  This  allows us to construct a phase diagram which  delimits the mechanical stability of capsids for different ionic strengths and shell surface charges.  
	\end{abstract}
	
	\maketitle
	\section{Introduction}

	There are many examples in nature of spontaneous self-assembly of nanoparticles that give rise to large-scale layers of closed sheet-like structures such as cages, membranes, connecting channels \cite{Mun12}, cavities, and closed pores. These topological structures have a convenient ability to enclose and hold part of the surrounding material inside them \cite{karp,Deserno,Cos03,Jas15,Hu15,Ser16}. The molecular forces that drive the self-assembly and control the stability of such enclosing surfaces are very sensitive to specific environmental conditions such as solvent quality, temperature, concentration, pH, shape and size of small constituents ({\it e. g.}, building blocks), the presence of 
	dissolved charged groups, among others \cite{Mar02,Mur09,Rij13,Matt13,Wu14,Bol16}. The proper understanding and control of the mechanisms that trigger the self-assembly of basic nano-sized elements into specific geometry -- as well as their stability under a wide range of physically attainable conditions -- represents a rather challenging task and an active area of investigation with tremendous practical implications~\cite{Deserno,Des09,Nog13,Per15,Bol16,Ser16}. A promising application relies on the encapsulation and release of nanoparticles at controlled targets \cite{Kho19}, such as in the engineering of active transport agents in drug-delivery applications \cite{Hol05,Fat14,Tha18,Pan18,Dad19}.   A number of relevant physical and biological processes have as a byproduct a partial formation of closed shells of different typologies, sizes, and internal structure~\cite{Gel08,Des09,Schm17,Cha19,Ger19}. In most of these situations, the assembled shell-like ``envelope'' is endowed with a very clear functionality, being naturally engineered to perform some specific biological task, such as to protect its internal content from an external (usually hostile) environment or to deliver and inject it at specific locations. Examples of such biological functional objects are numerous, ranging from cell membranes to liposome vesicles~\cite{Lev04,Idi04,Levin04}, lipid bilayers~\cite{Coo06,Des15}, phospholipid membranes, and viral capsids~\cite{Mate13}. The resulting closed sheets are typically semi-permeable, allowing for a selective control of the in- and outward material flux across their interfaces, thus acting as an interface which effectively shields its own content from undesired external stimuli.

	The basic features that control the stability of a nanoshell are the osmotic equilibrium over its surface -- which implies the equality of the various chemical potentials across this interface \cite{Loz96,Lev02,Des09,Lozada} -- and the elastic properties that dictates the surface ability to deform in response to the resulting osmotic stresses~\cite{Zan05,Ngu05,Ngu06,Mic06,Hu13,Kri16}. Despite their diversity in shape and deformability~\cite{Nor06}, a number of  biological membranes can be fairly well described as spherical-like, semi-permeable membranes suspended in an aqueous medium. In such a case, the main driving forces controlling the self-assembly and stability are hydrophobic (hydrophilic), elastic, and electrostatic in nature~\cite{Mou87,Jav19}. Along these lines, a great deal of work has been recently devoted to highlight the interplay between these interactions in the formation and equilibrium of biological-inspired nanoshells~\cite{Sil12,Her15,Sho16,Kov17,Sun18,Anv18,Reg19,Xia19}.

	One of the most versatile and robust among the biological-engineered shells are the capsids that make up the core structure of a virus. These so-called viral capsids are composed of small groups of coat proteins which self-assemble into small basic subunits, the {\it capsomers}. These building blocks are collectively combined to form an envelope that protects and transports the viral genetic material \cite{Almendral,Roo07,Joh10,Mat13,Com14,Bru16}. Viral capsids display a wide range of topologies~\cite{Ngu05,Sil06,Che07,Mar12,Abr12,Lov13,Pana18}, and can be further coated with membranes or protein sheets~\cite{Suc06,Sin08,Mat13,Cos14,Mil15}. The key aspects that describe the rich topology of these objects have been long ago rationalized in the seminal work of Clasper and Klug~\cite{Cas62,Sil06}. Particularly important is the remarkable mechanical versatility of viral capsids~\cite{Lov13}, which allows them to successfully protect, transport and release the viral genetic code through the (often insecure) cell environment. The accomplishment of these tasks requires the ability to sustain astonishing mechanical stresses and indentations at specific surface sites without loosing the elastic character~\cite{Zan05,Roo07,Sil09}, and yet keeping overall shell stability against disassembly of their constituents capsomers.

	Electrostatic forces are one of the key mechanisms that controls formation and stability of viral capsids, being able to either favor or inhibit capsid formation. Usually the coat proteins bear a net charge  due to ionic dissociation. In the case of empty capsids, the surface formation thus requires that the binding force be high enough to overcome the electrostatic repulsion between neighboring capsomers~\cite{Cer02,Sil08}. In many situations where the capsid is wrapped around a nucleation core of opposite charge ({\it e. g.}, the viral  genetic code or synthetic materials such as gold nanoparticles~\cite{Loo06,Per14,Jav19} or anionic proteins~\cite{Jas14,Bru16}), electrostatic interactions are the main driving force for surface self-assembly~\cite{Hag09,Jav13,Jas14,Jav19}. Since both strength and range of these interactions can be experimentally tunned {\it via} addition of ionizable groups at the capsomers or into the bulk solution~\cite{Lev02,Mes09,Fre10}, it is of fundamental importance to understand how the equilibrium and stability of biological  shells is influenced by ionic interactions~\cite{Sil08}. The aim of this work is to address this question by investigating the osmotic equilibrium and the elastic stability of shells over a wide range of ionic strengths and net surface charges.  To this end, a Density Functional Theory (DFT) is applied which allows one to accurately account for the ionic osmotic equilibrium across an infinitely thin spherical charged surface that represents a coarse grained description of an icosahedral, quasi-spherical, shell. The applied DFT accurately incorporates size and electrostatic correlations into the calculation of ionic free energies~\cite{Col16}. We then combine the results of the DFT with a continuum model of shell elasticity, allowing us to calculate the osmotic stresses across the interface and the equilibrium shell size. Conditions of mechanical stability against rupture and pore nucleation are then analyzed in a manner similar to the classical nucleation theory~\cite{Zan06}.

	The remainder of the paper is organized as follows. In section II, the basic model employed in the description of biologically relevant shells is described. Next, the theoretical approaches applied in the context of this model system to obtain equilibrium ionic properties -- namely the Density Functional Theory and Monte Carlo simulations -- are outlined in Sections III and IV, respectively. The reader already familiar with these approaches can skip these sections and head directly to Section V, where the main Results are shown and discussed in detail.
	Finally, concluding remarks and future perspectives are drawn in Section VI.

	\section{Model System}
	
\begin{figure}
		\centering
		\includegraphics[width=6.5cm,height=3.7cm]{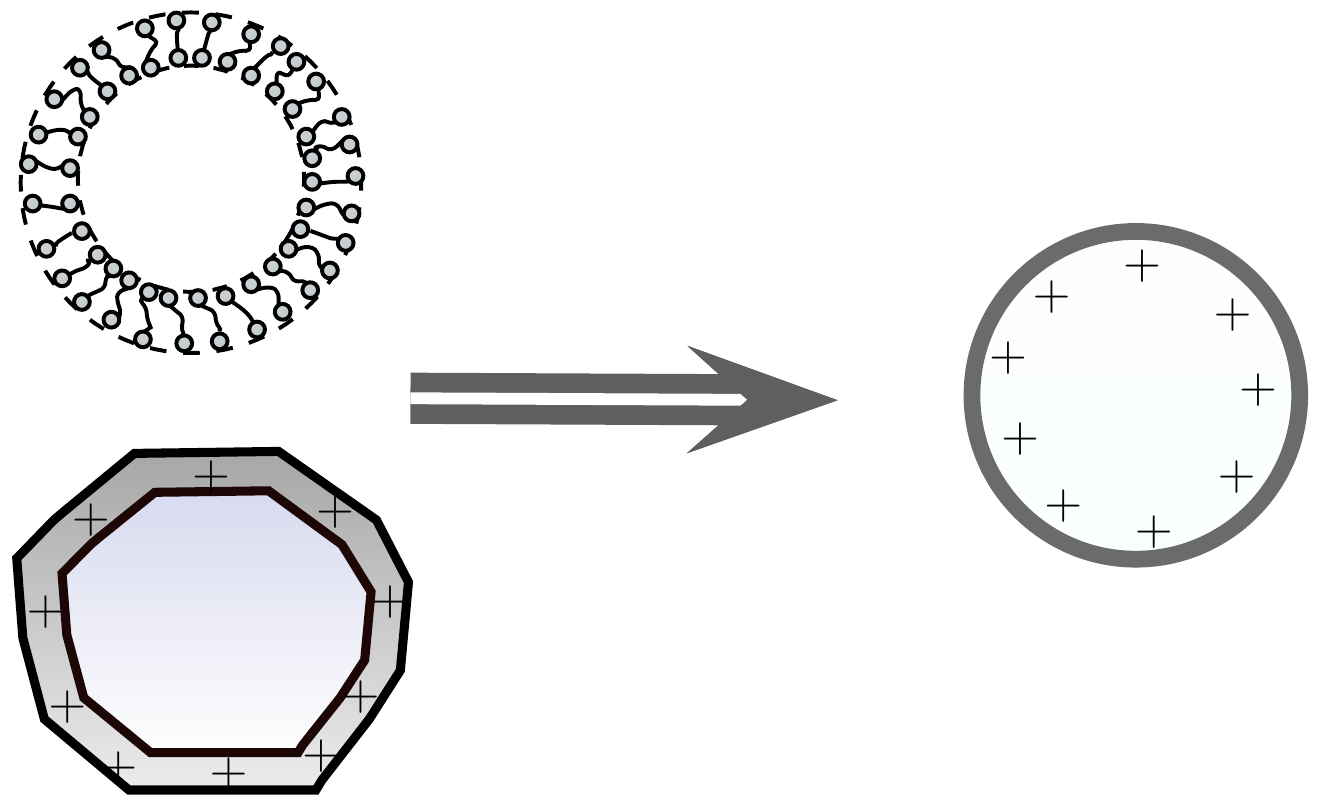}
		\caption{A sketch of coarse graining of biological shells. A biological nanoshell such as a liposome vesicle or an irregular icosahedral capsid is mapped onto a thin, structureless spherical shell bearing a net positive charge uniformly distributed over its surface. The shell is permeable to solvent molecules as well as to ions present inside the solution.}
		\label{fig:fig1}
	\end{figure}

	Despite the diversity of shapes and sizes of self-assembled biological capsids, we shall here focus, for the sake of simplicity, on the most common case of nearly-spherical objects. We consider an isolated shell in osmotic equilibrium with an aqueous solution containing a dissolved 1:1 symmetric electrolyte of bulk concentration $c_s$. We apply a coarse graining description to the capsid that averages out the fine local structures of the assembled capsomers, and results in a structureless spherical shell of radius $R$ and vanishing thickness, as depicted in Fig. \ref{fig:fig1}. The shell is further assigned with a homogeneous positive net charge $Zq$, representing the charges on the capsomers that make up the viral capsid or any prototypical spherical membrane  (here $q$ is the charge of a proton). Here we shall assume fixed surface charges, so that charge regulation effects are not taken into account at this level of approximation. Although such mechanisms are known to have a non-trivial influence on both the ionic diffusion and structure across the double layers~\cite{Nin71,Tre16,Mar17,Pod18,Smi18,Fry19,Bak19}, our current aim is rather to highlight the effects of salt addition on the mechanical shell properties at {\it fixed} surface charges. Ions are free to move throughout the system, and are allowed to diffuse through the semi-permeable membrane, just like the solvent molecules. In practice, the shell is  placed at the center of a much bigger spherical Wigner-Seitz cell, which is large enough to guarantee that the ions achieve a uniform bulk concentration $c_s$ far away from the charged shell. It is worthwhile to note that our model system takes into account only one isolated shell in equilibrium with a bulk electrolyte. Contrary to the usual WS cell approaches in which the WS cell size is chosen so as to account for the overall concentration of colloidal particles, here the cell size is arbitrarily large -- corresponding to the infinite dilution limit. The solvent is modeled as a structureless background of dielectric constant $\varepsilon\approx 80$, representing the surrounding aqueous environment. Throughout this work, we shall consider monovalent ions of hydrated radii $r=2$~\AA. 
	
	Although the capsomers in a viral capsid are usually assembled around a charged nucleus (composed by a close-packed genetic material), we shall here focus on the situation of {\it empty} capsids, which are also abundant in nature. Such capsids are able to spontaneously self-assemble without the aid of a nucleus and, therefore, contain no genetic material in their interior. The important question of capsids filled with an oppositely charged genetic material will be addressed elsewhere.     
	
	It might be argued that the proposed representation of intrinsically complex viral capsids as simple permeable spherical thin shells is a rather crude approximation in view of the actual heterogeneity in shape and local arrangements that these objects can display. However, our main goal here is to highlight the effects of electrostatics and excluded volume interactions, as well as their interplay with the overall mechanical properties of the shells.  Taking into account specific structural details of a given nanoshell would obscure the main physical mechanisms we aim to elucidate. Furthermore, as mentioned in the introduction, the variety of possible morphologies that these shells can assume is so large that a general microscopic model is simply unfeasible. In this sense, a simplified approach aimed at a wide class of systems should be very useful to shed light on the essential features that ultimately control much of the behavior of such systems.

	
	\section{Density Functional Theory}
	
	In order to calculate the ionic density profiles and the resulting equilibrium properties of the model system discussed above we employ a Density Functional Theory.  This approach is known to be very accurate for ionic systems in both strong and weak coupling regimes -- whereby effects from ionic correlations at the double layer are dominant and negligible~\cite{Sam11,Sam16_2,Sam16}, respectively -- and can also efficiently account for the excluded volume effects,  which become important in the interior of a capsid. The shell is taken to have a fixed charge, providing an external  field in the presence of which the mobile ions can freely diffuse. Placing a charged shell of radius $R$ and uniform charge $Zq$ at the origin of a coordinate system, the corresponding (dimensionless) electrostatic potential $\phi(\br)=\beta q \varphi(\br)$ (where $\beta=k_BT$ is the thermal energy, with $k_B$ denoting the Boltzmann constant and $T$ the temperature) produced by the capsid on an ion at position $\br$ is
	\begin{eqnarray}
	\phi(\br)  =  
	\begin{cases}
	&\dfrac{Z\lambda_B}{R},\qquad r\le R\vspace{0.25cm}\\
	&\dfrac{Z\lambda_B}{r},\qquad r>R.
	\end{cases},
	\label{shell_pot1}
	\end{eqnarray}
	where $\lambda_B=\beta q^2/\varepsilon$ is the Bjerrum length, which characterizes the typical length scale at which electrostatic and thermal contributions are similar in magnitude. We explicitly consider an aqueous system at room temperature, such that $\lambda_B\approx 7.2$~\AA. Apart from the above electrostatic interaction, the shell also provides a hard-core repulsion that avoids overlapping of the surrounding mobile ions of type $i$:
	
	\begin{eqnarray}
	\phi_i^{hc}(\br)  =  
	\begin{cases}
	&\infty, \qquad{\text if} \qquad r\le |R-a_i|,\vspace{0.25cm}\\
	&0\qquad {\text otherwise}.
	\end{cases},
	\label{shell_pot2}
	\end{eqnarray}
	
	Notice that the above potential is also normalized by the thermal energy $k_BT$. According to the classical DFT, the equilibrium distributions in the presence of an applied external field will correspond to the minimum of the free-energy functional $\Omega$, defined as~\cite{Hansen}
	\begin{equation}
	\beta\Omega=\sum_{i}\int\rho_{i}(\br)[\log(\Lambda^3\rho_i(\br))-1]d\br+\sum_i\int\left[z_i \phi (\br)+\phi_i^{hc}(\br)-\beta\mu_i\right]\rho_i(\br)d\br+\beta\mF^{ex},
	\label{FE1}
	\end{equation}
	where $\rho_i(\br)$ is the equilibrium density profile of ions of type $i$  with charge magnitude $z_i=\pm 1$ (normalized by the proton charge $q$), $\Lambda$ is the de-Broglie wavelength. In equilibrium, the above functional should match the appropriate thermodynamic free-energy. The first term on the right-hand side of this expression represents the entropic contributions from the mobile ions, while the second term comprises the interactions with the external field, as well as the coupling with the chemical potentials $\mu_i$, which ensure the condition of fixed bulk concentrations $c_i$ away from the charged surface. The last term, $\beta\mF^{ex}$, represents the intrinsic free-energy contributions resulting from inter-particle interactions. Defining the corresponding excess chemical potential $\mu^{ex}_i(\br)$ as the functional derivative  $\mu^{ex}_i=\dfrac{\delta \mF^{ex}}{\delta\rho_i(\br)}$, an application of the standard Euler-Lagrange stationary condition
	\begin{equation}
	\dfrac{\delta \beta\Omega}{\delta\rho_i(\br)}=0,
	\end{equation}
	to the above functional, Eq. (\ref{FE1}), yields the following equilibrium ionic profiles:
	\begin{equation}
	\rho_i(\br)=\bar{\rho}_i\exp\left[-\beta\mu_i^{ex}(\br)-z_i\phi(\br)-\phi^{hs}_i(\br)\right],
	\label{pro1}
	\end{equation}
	where $\bar{\rho}_i\equiv \dfrac{e^{\beta\mu_i}}{\Lambda^3}$. The Lagrange Multipliers $\mu_i$ are to be calculated in such a way as to satisfy the asymptotic condition $\rho_i(\br)=c_i$, {\it i. e.}, fixed bulk salt concentration. The excess free-energy in Eq. (\ref{FE1}) can be written as a superposition of decoupled electrostatic and ionic hard-sphere contributions, $\mF^{ex}(\br)=\mF^{el}(\br)+\mF^{hs}(\br)$. Accordingly, the excess chemical potential will be divided into (ionic) electrostatic and finite-size contributions, $\mu^{ex}_i(\br)=\mu^{el}_i(\br)+\mu^{hs}_i(\br)$. Once these quantities are known, the equilibrium ionic profiles and resulting ionic contributions to the total free-energy can be readily computed. Unfortunately, an accurate calculation of electrostatic and size contributions is not always achievable, even though different levels of approximation can always be employed~\cite{Hansen,Lev02}. The electrostatic ionic interactions can be further split into mean-field and correlational contributions, $\mF^{el}=\mF^{mf}+\mF^{cor}$. The former can be written in a closed form as~\cite{Lev02,Yan15} 
	\begin{equation}
	\beta\mF^{mf}=\dfrac{\lambda_B}{2}\sum_{ij}z_iz_j\int\dfrac{\rho_i(\br)\rho_j(\br')}{|\br-\br'|}d\br d\br'.
	\label{F_mf}
	\end{equation}
	This contribution clearly represents the (mean-field) electrostatic energy due to the mutual Coulomb interactions among mobile ions. When further combined with the electrostatic ionic interactions with the fixed charged shell in Eqs. (\ref{shell_pot1}), (\ref{shell_pot2}) and (\ref{FE1}), these contributions provide the overall ionic electrostatic energy at the mean field level. The correlational free-energy $\mF^{cor}$ can thus be interpreted as additional corrections in the ionic mean-field electrostatic energy which incorporate ionic electrostatic correlations. Note that the chemical potential corresponding to the above mean-field contribution is simply $\mu^{mf}_i(\br)=z_i\psi_{ion}(\br)$, where $\psi_{ion}(\br)$ is the  electrostatic potential (normalized by a factor $\beta q$) resulting from the inhomogeneous ionic distributions:
	\begin{equation}
	\psi_{ion}(\br)=\lambda_B\sum_iz_i\int\dfrac{\rho_i(\br')}{|\br-\br'|}d\br'.
	\label{pot_ion}
	\end{equation}
	We now define the {\it total} mean electrostatic potential as
	\begin{equation}
	\psi(\br)=\psi_{ion}(\br)+\sum_jz_j\int\rho_j(\br)\phi(\br)d\br.
	\label{pot_mf}
	\end{equation}
	Putting together all these results for the excess chemical potential into the equilibrium Euler-Lagrange condition Eq. (\ref{pro1}) yields the following simplified relation for the ionic density profiles around the charged shell:
	\begin{equation}
	\rho_i(\br)=\bar{\rho}_i\exp\left[-z_i\psi(\br)-\beta\mu_i^{cor}(\br)-\phi^{hs}_i(\br)-\beta\mu_i^{hs}(\br)\right].
	\label{EL}
	\end{equation}
	The Eq. (\ref{EL}) can now be combined with the Poisson equation to self-consistently determine the electrostatic potential in both the interior and the exterior of the capsid.  In practice, due to spherical symmetry we use the Gauss law to calculate the mean-field potential $\psi(\br)$ produced by the ions and thus close all the equations.  The traditional Poisson-Boltzmann (PB) theory is recovered when both size and electrostatic correlations are neglected~\cite{Lev02} ($\mu_i^{cor}=\mu_i^{hs}=0$). For computing these contributions beyond the mean-field level, we shall here invoke a second-order bulk functional expansion approximation for the electrostatic correlations~\cite{Ros93,Jia14,Yan15,Col16}, while describing finite size effects in the framework of the accurate Fundamental Measure Theory (FMT). 
	
	The FMT was designed to account for the hard-core effects in different geometries~\cite{Ros89,Ros90,Rot10}. In the FMT, the excess hard-sphere free-energy $\mF^{hs}$ is constructed  as a functional of a proper set of weighted densities $n_{\alpha}(\br)$:
	\begin{equation}
	\beta\mF^{hs}[n_{\alpha}(\br)]=\int\beta\Phi(n_{\alpha}(\br))d\br,
	\label{F_hs}
	\end{equation}
	where $\Phi(n_{\alpha}(\br))$ is a {\it local} free-energy density which only depends on the weighted densities (here labeled with the index $\alpha$) evaluated at a single point. These quantities are constructed {\it via} convolutions of the actual profiles with suitable normalized weight functions which can be scalar, vector, or even tensor in nature, and whose typical length scale depends on the particle size~\cite{Rot02,Rot10,Dav16}. Here we use the six traditional weight functions introduced by Rosenfeld in his pioneering work on FMT. The energy density $\Phi(n_{\alpha}(\br))$ is not unique, and is constructed so as to fulfill some requirements that control the accuracy of the underlying functional. Here we make use of the so-called White-Bear (WB) functional, known to be very accurate for describing the phase diagram of polydisperse hard-spheres in a wide range of relative concentrations. The corresponding energy density reads~\cite{Rot02,Rot10}:
	\begin{equation}
	\beta\Phi(n_{\alpha}(\br))=(n_2^3-3|\bn_2|^2n_2)\dfrac{(1-n_3)^2\log(1-n_3)+n_3}{36\pi n_3^2(1-n_3)^2}-\dfrac{(\bn_1\cdot\bn_2-n_1n_2)}{1-n_3}-n_0\log(1-n_3).
	\label{WB}
	\end{equation}
	
	In the Supplementary Information, we provide explicit relations for the FMT weighted densities and the resulting chemical potentials for systems with spherical geometry. Finally, the electrostatic correlational contribution $\mF^{cor}$ is calculated considering the following second-order functional expansion of this quantity in terms of the reference homogeneous system whose concentrations are the same as of the bulk~\cite{Zho05,Col16,Col17}:
	\begin{equation}
	\beta\mF^{cor}[\rho_i(\br)]\approx\beta\mF^{cor}[c_i]+\sum_i\beta\mu_i^{bulk}\int\delta\rho_i(\br)d\br-\dfrac{1}{2}\sum_{ij}\int c_{ij}(|\br-\br'|)\delta\rho_i(\br)\delta\rho_j(\br')d\br d\br'.
	\label{F_cor}
	\end{equation}  
	Here, $\mu_i^{bulk}=\dfrac{\partial \mF^{cor}}{\partial c_i}$ are homogeneous correlational chemical potentials of the corresponding bulk system, $\delta\rho_i(\br)=\rho_i(\br)-c_i$ are the local density deviations with respect to their bulk counterparts (and thus vanish very quickly as we move away from the charged membrane), and $c_{ij}^{bulk}(|\br-\br'|)=-\dfrac{\delta^2\beta\mF^{cor}}{\delta\rho_j(\br')\delta\rho_i(\br)}$ are the so-called direct correlation functions for the correlational free-energy, evaluated in the limit of a homogeneous electrolyte with bulk concentrations. The advantage of this approach is that this quantity can be readily evaluated in the framework of the Ornstein-Zernike (OZ) integral equations for homogeneous systems, with varying levels of accuracy. Once the direct pair correlations for the underlying homogeneous electrolyte are known, the correlational contributions to the electrostatic energy can be directly computed from Eq. (\ref{mu_cor}). The corresponding chemical potentials are
	\begin{equation}
	\beta\mu_{i}^{cor}(\br)=\beta\mu_i^{bulk}-\sum_{j}\int c_{ij}^{bulk}(|\br-\br'|)\delta\rho_j(\br')d\br'.
	\label{mu_cor}
	\end{equation}  
	Notice that the homogeneous bulk chemical potentials can be in practice incorporated into the Lagrange Multipliers $\mu_i$, and therefore need not to be explicitly taken into account in the present treatment~\cite{Col16}. Eqs. (\ref{F_cor}) and (\ref{mu_cor}) become increasingly more accurate as the density profiles deviations from the homogeneous bulk values become either small in magnitude or very localized. In order to calculate the direct pair correlations, the OZ equations~\cite{Hansen} 
	\begin{equation}
	h_{ij}(\br)=c_{ij}(\br)+\sum_k \rho_k\int h_{ik}(\br')c_{kj}(|\br-\br'|)d\br',
	\label{OZ}
	\end{equation}   
	are numerically solved. Here, $h_{ij}(\br)$ and $c_{ij}(\br)$ are, respectively, the total and direct correlation functions for the homogeneous electrolyte of bulk concentrations $\rho_k=c_k$ (from here on we omit the superscription {\it bulk}). An additional closure relation between direct and total bulk correlations is necessary for completeness, which we here set to be the well-known hyppernetted-chain (HNC) approximation,
	\begin{equation}
	h_{ij}(\br)=\exp[-\beta u_{ij}+h_{ij}(\br)-c_{ij}(\br)]-1,
	\label{OZ}
	\end{equation}
	which is quite accurate up to very high electrostatic couplings of the underlying bulk electrolyte~\cite{Col16}. The pair interactions $u_{ij}(\br)$ comprise both excluded volume and the electrostatic interactions. However, since the pair correlations in Eq. (\ref{mu_cor}) refer only to the {\it electrostatic correlations} (ionic size effects are accounted for {\it via} the FMT approach), the hard-core contributions have to be removed from the overall correlations before application of Eq. (\ref{mu_cor}). To this end, the OZ equation has to be solved for both {\it charged} and {\it uncharged} ionic bulk systems separately. The electrostatic direct correlations are then obtained by simply subtracting the calculated size contributions (for the uncharged system) from the total direct correlation function, before plugging them into Eqs. (\ref{F_cor}) and (\ref{mu_cor}). 
	
	After calculating the mean-field, hardcore, and correlational free-energy contributions $\mF^{mf}$, $\mF^{hs}$ and $\mF^{cor}$ from Eqs. (\ref{F_mf}), (\ref{F_cor}) and (\ref{F_hs}), respectively, the excess free-energy follows directly from the combination $\mF^{ex}=\mF^{mf}+\mF^{cor}+\mF^{hs}$ , along with the total ionic free-energy, Eq. (\ref{FE1}). Application of the Euler-Lagrange condition (\ref{pro1}) then provides the equilibrium ionic profiles, from which all the equilibrium properties of the ionic system subjected to the shell field can be inferred.

	\section{Monte Carlo Simulations}
	
	Apart from the above described DFT approach, we will also make use of equilibrium Monte Carlo (MC) simulations to further access ionic equilibrium properties of the proposed model system and to check the accuracy of the DFT. To this end, we will use a Primitive Model (PM) electrolyte in equilibrium with a charged porous shell. The (initially empty) cavity is
	represented as a hollow sphere of radius $R$. Water is modeled
	as a uniform dielectric of permittivity $\varepsilon\approx80$. The system is at room temperature, so that the
	Bjerrum length is set to $\lambda_B=7.2$~\AA. The simulations are performed inside a spherical Wigner-Seitz (WS) cell of radius $R_c\gg R$, with a uniformly charged shell of charge $Zq$ placed at the center. The cell also contains $N = Z$ dissolved counterions, each of diameter $d_i=2r_i=4$~\AA.
	The interaction between the charged shell and an ion of charge $z_i q$ ($i=\pm$) at position $\br_i$ is $\beta U_{si}(\br_i)=z_i\phi(r_i)$, where $\phi(r)$ is the electrostatic shell potential in Eq. (\ref{shell_pot1}). Ions are free to move all over the WS cell, and interact with each other thorough Coulomb and hardcore potentials, $\beta u(\br_i,\br_j)=\beta u_{c}(\br_i,\br_j)+\beta u_{hs}(\br_i,\br_j)$, where $\beta u_{c}(\br_i,\br_j)=\lambda_B z_i z_j/|\br_i-\br_j|$ and $\beta u_{hs}(\br_i,\br_j)=\infty$ if $|\br_i-\br_j|\le d_{ij}=(d_i+d_j)/2$ and $u_{hs}(\br_i,\br_j)=0$ otherwise. The total ionic energy for a given configuration is
	\begin{equation}
	\begin{split}
	\beta U = \sum_{i} z_i\phi(\br_i) +\dfrac{1}{2}\sum_{\substack{ij \\ (i\neq j)}} \beta u(\br_i,\br_j),
	\end{split}
	\label{U}
	\end{equation}
	where the sums run over all the ions inside the WS cell. In the simulation the overlap between ions and the capsid shell is not permitted, however, the ions can ``jump" across the shell.
	We use the Eq. (\ref{U}) in a typical Metropolis algorithm, with $10^7$ MC steps for equilibration
	and $10^5$ steps for production. The ionic density profiles are obtained by dividing the WS cell into concentric spherical bins and counting the average number of particles in each bin for all uncorrelated configurations~\cite{Allen}.
	
	Apart from the $Z$ dissociated counterions, the WS cell can also contain 1:1 electrolyte of average concentration $\bar{\rho}_s=N_s/V$, where $N_s$ is the number of dissociated pairs, and $V$ the cell volume. 
	Chemical equilibrium with the bulk electrolyte is established indirectly by changing the overall concentration of salt inside the simulation cell, until the ionic profiles achieve the desired bulk values far away from the capsid surface. To this end, the cell radius $R_c$ has to be chosen large enough to avoid spurious finite size effects. Notice that the simulations are performed for an isolated capsid, so that use of special techniques to handle long-range Coulomb interactions in the presence of infinite replicas to account for finite capsid concentration is not necessary. Moreover, image charges are absent due to the uniformity of the dielectric constant throughout the system.
	
	\section{Results and discussions}

	We will now employ the theoretical tools outlined above to investigate many equilibrium properties of empty capsids in an electrolyte solution. We will first analyze the ionic equilibrium properties, such as the ionic density profiles and the osmotic pressure built-up across the interface. After this we will address the question of the mechanical stability of the capsid, which depends on its ability to withstand the osmotic stress imposed by the environmental conditions ({\it e. g.}, the salt concentration of the solution).

	\subsection{Ionic Profiles}
	
\begin{figure}[h!]
		\centering
		\includegraphics[width=7.25cm,height=5cm]{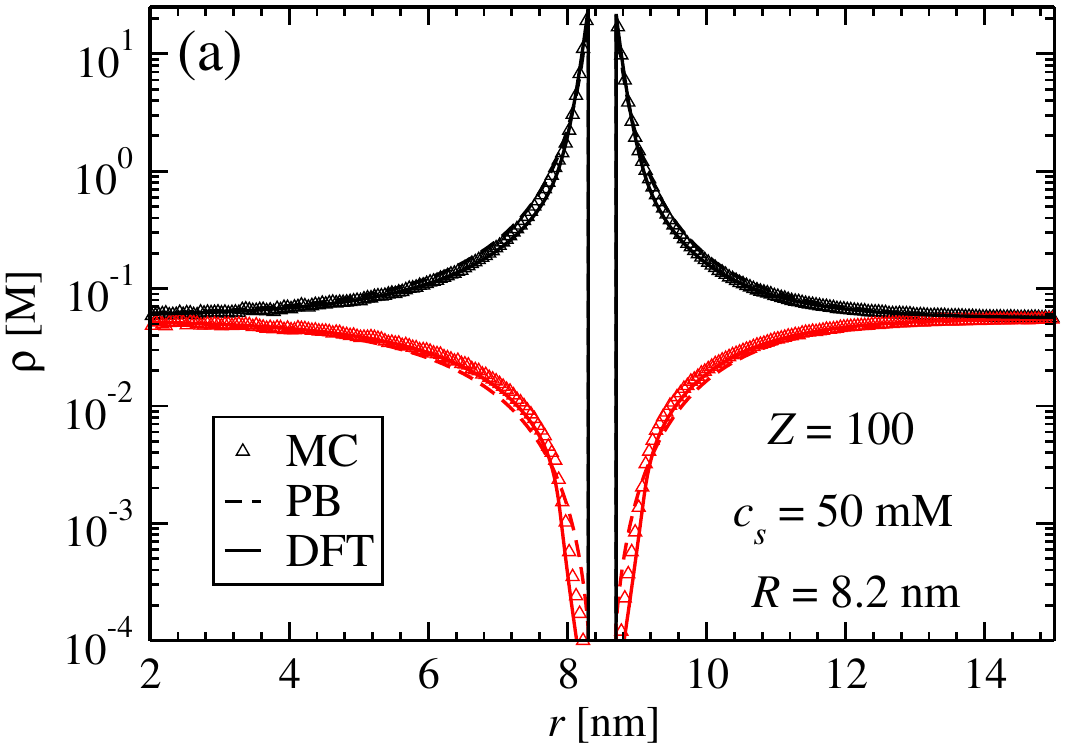}
		\includegraphics[width=7.25cm,height=5cm]{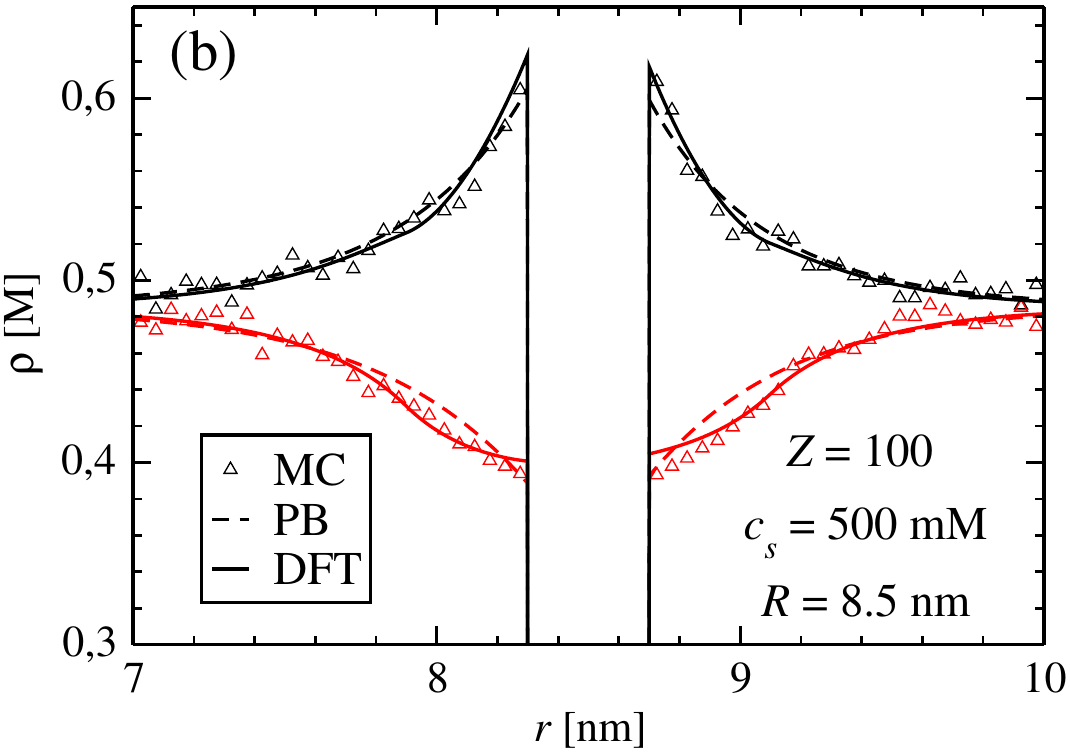}\\ \hspace{0.2cm}
		\includegraphics[width=7.25cm,height=5cm]{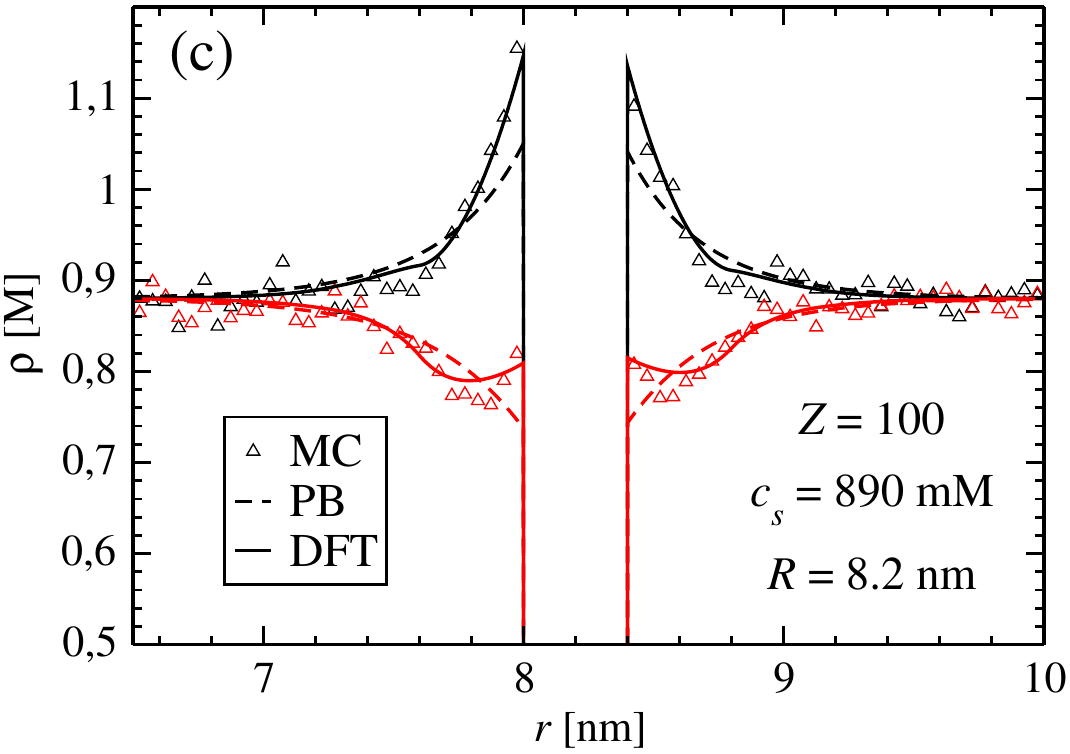}
		\includegraphics[width=7.25cm,height=5cm]{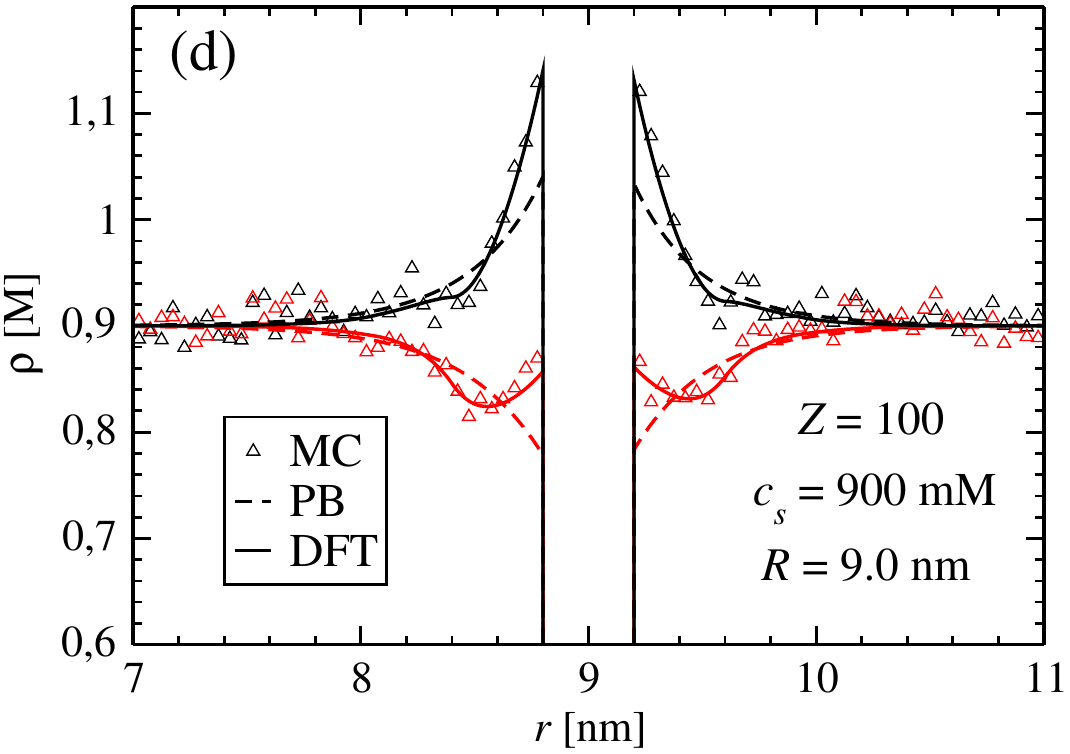}
		\caption{Ionic profiles in the vicinity of a charged shell of varying charge and size in the case of low (a) and high (b), (c) and (d) ionic strengths. The black curves are the counterion profiles, while red curves correspond to coion distributions. Symbols are MC results, dashed lines represent the mean-field PB results, whereas the solid curves are profiles obtained from the DFT approach.}
		\label{fig:fig2}
	\end{figure}

	As discussed above, ions are allowed to penetrate into the shell, although overlapping is always avoided through the hardcore repulsion. As a result, an electric double layers will be developed at both the internal and the external shell surfaces, with counterions on average located closer to the surface and coions farther from it. This trend is clearly observed in Fig. \ref{fig:fig2}, in which the ionic density profiles across the shell for different ionic strengths and various shell radii $R$ and charge $Z$ are shown.  We note that the infinitely thin layer can be viewed as a hard spherical wall carrying a uniform surface charge density $\sigma/2$ on each of its faces, where $\sigma=Zq/4\pi R^2$ is the overall shell charge density. Since the radius of curvature of the nanoshells are usually much larger than the range of averaged electrostatic interactions (which is measured by the inverse Debye screening length $\kappa^{-1}=(8\pi\lambda_B c_s)^{-1/2}$), the electrostatic potential will be effectively screened as one moves away from the surface. As a consequence, electroneutrality will be achieved both inside and outside the shell. This is reflected in a quick relaxation of the ionic profiles to their corresponding bulk values not only outside, but also in the interior of the charged cavity~\cite{Fre10}. The region of charge inhomogeneity is thus confined to the close vicinity of the shell, where thus all the relevant physics that controls shell stability should take place. If the double layers which are built up on both sides of the shell are fully symmetric, no osmotic stress will develop across the interface, and the surface will be unstretched. Furthermore, the larger the ionic strength, the narrower will be the double layer across the surface, as can be clearly observed in Fig. \ref{fig:fig2}. Although the surface charges are reasonably small in these cases, similar trends will be observed at larger surface charges as well (as long as the shell is empty). Fig. \ref{fig:fig2} also demonstrates the accuracy of the proposed DFT approach when compared against the MC data. To emphasize this point, comparisons with the mean-field Poisson-Boltzmann (PB) predictions are also displayed. At low ionic strengths ({\it i. e.}, small salt concentrations), the double layer is diffuse, and both PB and DFT approaches perform  equally well (see Fig. \ref{fig:fig2}a, where dashed and full curves become indistinguishable). In all these cases, the shell size is fixed, and has not been subject to the mechanical equilibrium condition to be described in what follows.

	
	As the salt concentration increases, the inhomogeneity range across the interface becomes increasingly small and ionic correlations resulting from packing start to play a relevant role. As a result, the ionic density profiles no longer display the monotonic behavior predicted by the mean-field approach and small layering-like structures start to appear due to strong positional correlations inside the double layers. This behavior is expected to be dominant in the case of high electrostatic couplings~\cite{bkh18}, but can also be observed in the present situation of low coupling and high salt concentrations. Clearly, such correlation effects tend to be more pronounced as the surface charge density of a shell increases -- as more and more counterions try to pack close as possible to the shell surface.  Contrary to the inherent failure of the mean-field approximation, the  DFT is able to quite well capture the fine structural features of the ionic double layer at ionic strengths where, as we will shortly see, the presence of salt may play a relevant role in dictating the shell mechanical stability. 
	
	\subsection{Osmotic Stress}
	
	In the proposed model system, the charged spherical shell can be also interpreted as a dividing surface which separates two coexisting electrolytes, namely the one confined to the shell interior and the other one in contact with a bulk salt reservoir. In this case, the applied DFT approach properly accounts for the chemical  equilibrium of all ionic species over the charged interface. In fact, the Euler-Lagrange condition, Eq. (\ref{EL}), is equivalent to the equality of the chemical potentials $\mu_i$ over the different phases. This does not, however, guarantee the {\it mechanical} equilibrium across the dividing interface. In the case of a mobile dividing surface, the location of the wall will relax to a position of vanishing osmotic pressure, thereby enforcing mechanical stability. If the surface has fixed position, an external pressure -- closely related to the so-called solvation force~\cite{Hansen} -- has to be exerted such as to ensure mechanical equilibrium. In the case of a stretching or deformable surface, the imposed mechanical stress will be balanced by a surface rearrangement ({\it i. e.}, volume or shape changes) that should guarantee an overall vanishing stress. In this case, the reaction to the applied stress comes at a cost of internal surface forces -- usually of elastic nature -- that balance the net stress, while keeping the surface integrity. Such force balance across an interface is a common feature of many soft-matter systems, ranging from cross-linked microgels~\cite{Col14,Den16} to polymer brushes~\cite{Ter17} and biological membranes~\cite{Des15}.  
	
	In the present situation, the unbalanced ionic distributions in the inner and outer shell layers will lead to the build-up of a radially symmetric osmotic stress across its surface. As a result, the shell will either shrink or expand in response to such compressing or stretching forces. On the other hand, if the resulting strains are too large, it might be energetically favorable for the shell to open up a stable pore or even to break apart to release the stress~\cite{Idi04}. The overall shell stability is dictated by a fine interplay between imposed osmotic stress and the internal membrane forces, which we shall here describe using a continuum elastic model.         
	
	In order to analyze the stability of the nano-sized elastic shell when its radius is allowed to change, the internal energy of the shell of a given size $R$ should be added to the ionic free-energy $\Omega$ in Eq. (\ref{FE1}). Within the proposed model, this energy will be comprised of an electrostatic self-energy $U^{self}$ of the $Z$ surface charges, in addition to an elastic stretching energy $U^{elas}$, which depends on the equilibrium (unstretched) size of the shell. The electrostatic self-energy of the shell can be readily calculated to be
	\begin{equation}
	\beta U^{self}=\dfrac{Z^2\lambda_B}{2R}.
	\label{Uself}
	\end{equation} 
	Notice that this contribution increases with the square of the shell charge. If the shell is highly charged, the electrostatic repulsion between its constituent building blocks will naturally produce a stretching force on the surface, which will tend to inflate the shell. In the absence of other contributions to stress, this outward stress should be balanced by a compressing elastic stress which acts as to restore an unstretched equilibrium state. Clearly, such restoring stress comes from the attractive forces that promote the self-assembly of the fundamental building-blocks into capsid. In the framework of a continuum elastic theory, the corresponding energy penalty for stretching a shell of original unstretched size $R_0$ is taken to be \cite{Eva87,Idi04}
	\begin{equation}
	\beta U^{elas}=\beta\dfrac{\kappa_s}{2}\dfrac{(A-A_0)^2}{A_0},
	\label{Uelas}
	\end{equation} 
	where $\kappa_s$ is the so-called {\it stretching modulus} of the nanoshell, $A_0=4\pi R_0^2$ is the unstretched surface area, and $A=4\pi R^2$ is the area after radial stretching/compression of the spherical shell. Notice the resemblance of the elastic energy with the one-dimensional Hook's Law. Similarly to that case, the above expression will be accurate as long as strains are not too large~\cite{Landau,Idi04}. It is also important to note that in the present model of a shell of vanishing thickness, no bending energy needs to be taken into account. 
	Upon stretching, the shells will be subject to an elastic restoring stress, 
	\begin{equation}
	\beta\Pi^{elas}=-\dfrac{1}{4\pi R^2}\dfrac{\partial \beta U^{elas}}{\partial R}=\dfrac{2\beta\kappa_s}{R}\left(1-\dfrac{R^2}{R_0^2}\right).
	\label{Pi_elas}
	\end{equation}
	On the other hand, the surface charge density will decrease upon shell stretching, so that the self-energy electrostatic contribution will be reduced, leading to an outward electrostatic surface stress. From Eq. (\ref{Uself}), one finds 
	\begin{equation}
	\beta\Pi^{self}=-\dfrac{1}{4\pi R^2}\dfrac{\partial\beta U^{self}}{\partial R}=\dfrac{\lambda_B Z^2}{8\pi R^4}. 
	\label{Pi_self}
	\end{equation}
	Note that this contributions scales with the square of the surface charge density. If the shell could be left isolated from any other external effects, these competing interactions would lead to a stretching of the surface until the condition of vanishing stress is achieved, $\Pi^{elas}+\Pi^{self}=0$. The presence of dissociated counterions and the possible addition of an electrolyte of concentration $c_s$ will obviously change this simple scenario, as the highly inhomogeneous ionic distributions across the shell will lead to an additional osmotic stress, which in turn depends on the fine asymmetry between internal and external double layer structures. In order to investigate the interplay between these contributions in the underlying mechanical equilibrium, we must also consider the effects of an ionic stress $\Pi^{ion}$ defined as 
	\begin{equation}
	\beta \Pi^{ion}=-\dfrac{1}{4\pi R^2}\dfrac{\partial\beta \Omega}{\partial R},         
	\label{Pi_ion}
	\end{equation}     
	where $\Omega$ is the total ionic free energy given by Eq. (\ref{FE1}). Notice that this quantity is nothing but the ionic osmotic pressure on the membrane, defined as the difference between internal and external radial pressures resulting from the shell interaction with the electrolyte. The osmotic stress $\Pi^{ion}$ is composed of two contributions: the electrostatic stress $\Pi^{elec}$ and the mechanical stress $\Pi^{mec}$. These quantities can be easily evaluated once the ionic profiles across the interface are known. Due to the Gauss's Law, the ionic electric field across the surface depends only on the net ionic charge lying {\it inside} the shell $Z_{in}$,
	\begin{equation}
	Z_{in}=4\pi\sum_i\int_{0}^{R}\rho_{i}(r)r^2dr.
	\label{Zin}
	\end{equation}
	Therefore, ions located outside the shell do not directly contribute to the electrostatic stress. Due to the strong adsorption of counterions at the inner capsid surface, $Z_{in}$ will be clearly negative. Since the ionic electric field on the surface has magnitude $E= Z_{in}q/\varepsilon R^2$, the resulting stress will be simply $\beta\Pi^{elec}=\lambda_B Z_{in} Z/4\pi R^4$. Notice that the electrostatic stress points either outward (inward) if the net internal charge $Z_{in}$ is positive (negative). Meanwhile, the mechanical osmotic pressure corresponds to the momentum transfer from ionic collisions with the surface on both sides of the charged wall, and can be readily evaluated as $\beta\Pi^{mech}=\rho(R_{-})-\rho(R_{+})$, where $\rho(R_{+})$ and $\rho(R_{-})$ represents the total ionic contact densities at the external and internal shell surfaces, respectively. A stronger adsorption at the internal (external) faces will result in an outward (inward) osmotic stress. Combining these results, the ionic osmotic pressure can be written in a simple and transparent form as
	\begin{equation}
	\beta \Pi^{ion}=\rho(R_{-})-\rho(R_{+})+\dfrac{\lambda_B Z_{in} Z}{4\pi R^4}.
	\label{Pi_ion}
	\end{equation}
	Alternatively, the above quantity can also be computed by numerically evaluating the derivative in Eq. (\ref{Pi_ion}), after the ionic free-energy for different shell sizes $\Omega(R)$ is computed. We confirm that both approaches provide the same results within the numerical accuracy. A detailed derivation of Eq. (\ref{Pi_ion}) based on the ionic distribution around the charged shell is provided in the Supplementary Information. While the  contact stress can be either positive or negative, the second contribution in Eq. (\ref{Pi_ion}) always leads to the shell compression. By combining the above ionic contribution with the electrostatic self-energy due to the dissociated charged ionic groups in the aqueous environment (Eq. (\ref{Pi_self})), the total osmotic stress $\Pi^{osm}=\Pi^{ion}+\Pi^{self}$ is obtained:
	\begin{equation}
	\beta\Pi^{osm}=\rho(R_{+})-\rho(R_{-})+\dfrac{\lambda_B Z(Z+2 Z_{in})}{8\pi R^4}.
	\label{Pi_osm}
	\end{equation}   
	
	The mechanical equilibrium condition for the shell membrane will then be fulfilled when the total  stress vanishes, {\it i. e.}, $\Pi^{total}=\beta \Pi^{osm}+\beta\Pi^{elas}=0$. Explicitly, this condition can be written as:
	\begin{equation}
	\rho(R_{+})-\rho(R_{-})+\dfrac{\lambda_B Z(Z+2 Z_{in})}{8 \pi R^4}=\dfrac{2\beta\kappa_s}{R}\left(\dfrac{R^2}{R_0^2}-1\right).
	\label{equi}
	\end{equation}
	The relation above contains all the relevant physical contributions that determine mechanical equilibrium across the shell membrane. Note that the quantities on the left-hand side depend on the particular environmental conditions and the ionic equilibrium across the membrane, while the term on the right-hand side depends solely on the material properties through the elastic parameter $\kappa_s$. A quite clear physical interpretation can be assigned to this relation: the externally imposed stress must be counterbalanced by the mechanical ability of the material to rearrange its size in order to sustain the applied tension. It is also important to notice that the quantity $\Pi^{osm}$ depends on the specific structural features of the double-layers that build-up across the shell. In most cases, charge neutrality is satisfied inside a big shell, so that $Z_{in}\approx -Z/2$ and the osmotic stress simplifies to $\Pi^{osm}\approx \rho(R_{+})-\rho(R_{-})$. In this case, the precise determination of ionic densities at close contact with the inner/outer walls plays a major role in determining the mechanical equilibrium properties across the electrolyte interface, as anticipated in some previous works~\cite{Mal15}.  
	
\begin{figure}[h!]
		\centering
		\includegraphics[width=4.95cm,height=4.2cm]{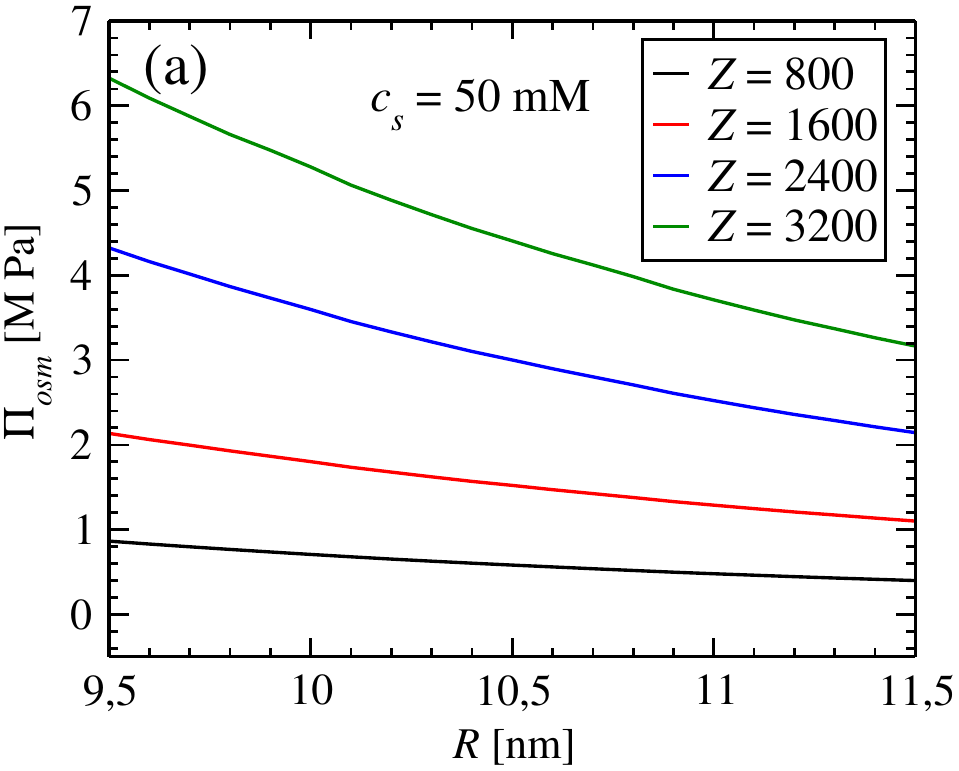}
		\includegraphics[width=4.95cm,height=4.2cm]{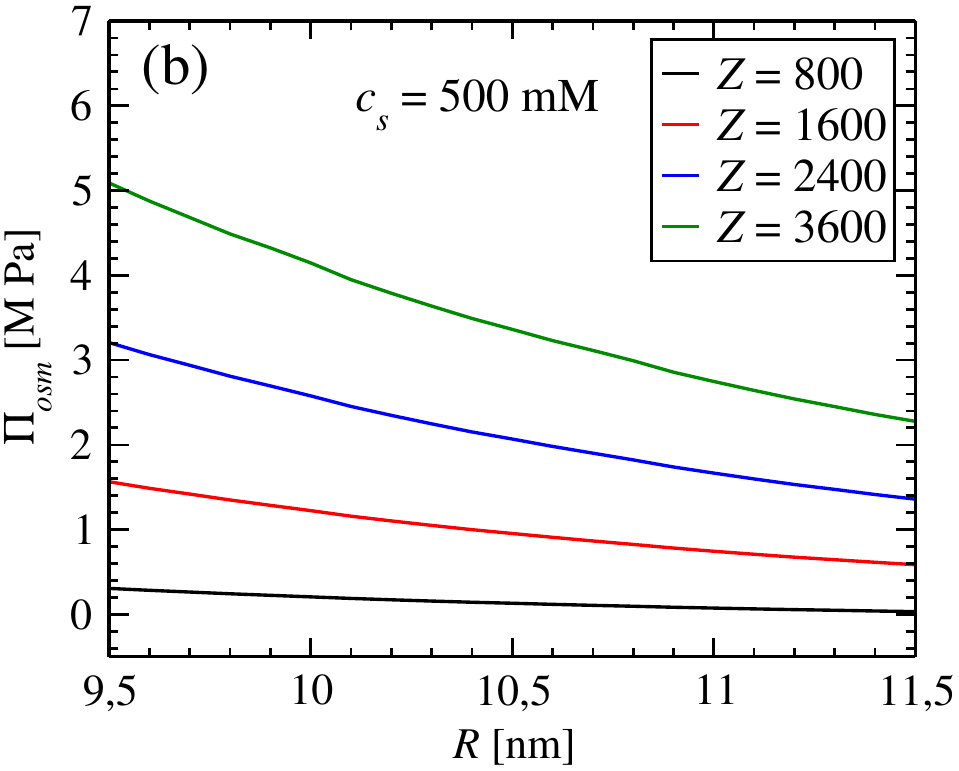}
		\includegraphics[width=4.95cm,height=4.2cm]{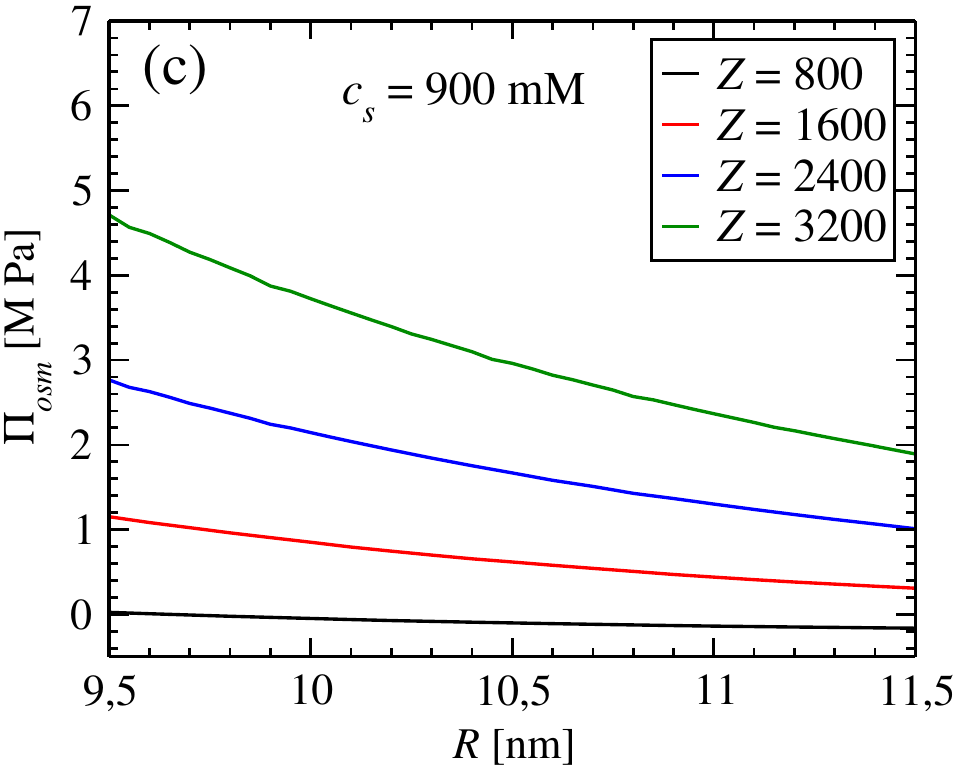}
		\caption{Osmotic stress $\Pi^{osm}$ (defined in Eq. (\ref{Pi_osm})) due to the combined electrostatic self-energy and ionic contributions as a function of the shell size in the range $9.5$~nm$\le R \le 11.5$~nm. The added electrolyte concentrations are $c_s=50$~mM (a), $c_s=500$~mM (b), and $c_s=900$~mM (c). The corresponding shell charges are displayed in the figures.}
		\label{fig:fig3}
	\end{figure}

	The behavior of $\Pi^{osm}$ (see Eq. (\ref{Pi_osm})) as a function of the shell deformations around the unstretched radius of $R_0=10$~nm is shown in Fig. \ref{fig:fig3} for different shell charges $Z$ and ionic bulk concentrations $c_s$. We first note that the shell charge has, as expected, a major effect on the osmotic stress. This behavior is most prominent for small shell sizes, while the effects of increasing shell charge become less important as the shell size grows. Clearly, this is a manifestation of the strong $\sim R^{-4}$ decay of the electrostatic contribution in (\ref{equi}). We can, therefore, identify two main regimes for the osmotic stress on a shell: a regime of electrostatic dominance at large surface charges and/or small shell sizes and a salt-dominated regime at high ionic strengths and small surface charge densities where the mechanical ionic stress $\rho(R_{-})-\rho(R_{+})$ becomes the major contribution. Indeed, it can be identified in Fig. (\ref{fig:fig3}) that addition of salt has a much stronger impact as the shell size increases and its charge decreases. We can also observe that the effects of salt addition are non-monotonic: at low ionic strengths, addition of salt has a much more pronounced effect on the osmotic stress. This can be traced back to the inability of a system to fully achieve internal charge neutrality for small ionic strengths. In particular, the osmotic stress at the smaller charge $Z=800$ and larger salt concentration $c_s=900$~mM becomes negative across the whole range of shell sizes (see Fig. \ref{fig:fig3}c).  This can lead to crumpling of the shell.   
	

\begin{figure}[h!]
		\centering
		\includegraphics[width=7.5cm,height=5.5cm]{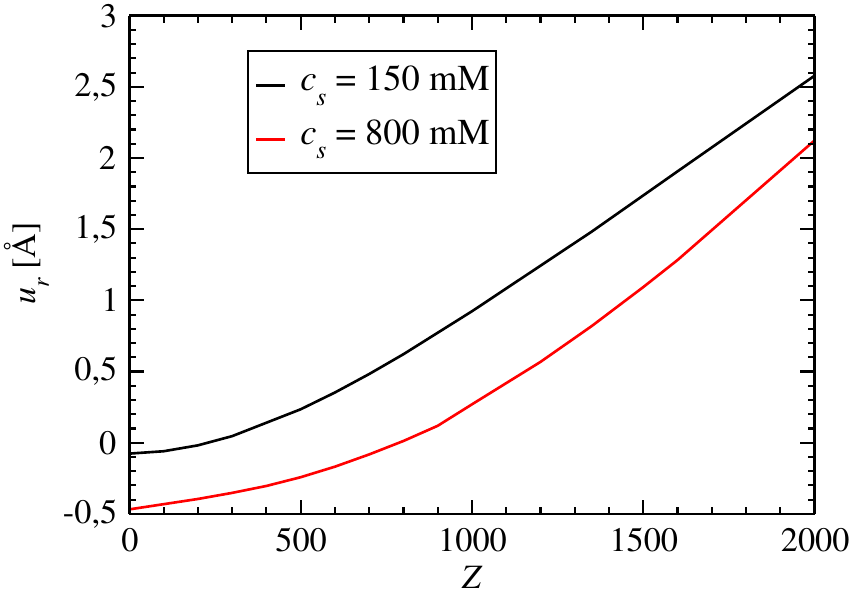}
		\caption{Radial strain $u_r=R-R_0$ as a function of the surface charge $Z$ for two representative salt concentrations $c_s=150$~mM (black line) and $c_s=800$~mM (black line). }
		\label{fig:fig4}
	\end{figure}

	We next analyze the shell's mechanical responses to the osmotic stress. Unfortunately a microscopic description is unfeasible in this case, since the deformations in response to the applied stress will depend on the particular properties of the shell material -- which we here incorporate into the elastic stretching modulus $\kappa_s$, in a coarse grained continuum approximation. First, we focus on a situation of a spherical capsid with equilibrium unstretched radius of $R_0=10$~nm.  Since all stresses are radially symmetric, the shell will keep its spherical shape and simply change its radius in response to the applied osmotic stress.  In what follows, we fix the elastic modulus of the shell at $\kappa_s=0.2$~J/m$^2\approx 48.6~ k_BT/$nm$^2$, which is a suitable value for empty viral capsids~\cite{Roo07}. By plugging this value into Eq. (\ref{equi}), the corresponding elastic stress upon stretching can be evaluated, and this equilibrium relation can be numerically solved to provide the equilibrium shell size $R$. Since the elastic stress is in this case $\Pi^{osm}\approx 40 (A/A_0-1)$~MPa, a close inspection of Fig. \ref{fig:fig3} allows one to conclude that the shell surface area will change at most by $\sim 15\%$, corresponding to a radius variation of order of a few percent with  respect to the relaxed shell. This rough estimate is confirmed in Fig. \ref{fig:fig4}, in which the radial strains $u_r\equiv R-R_0$ are shown as a function of the shell charge at two representative salt concentrations (here, $R_0=10$~nm). For vanishing surface charges, the electrostatic contributions are absent, and the ionic contact stress is negative, leading to a compression of the shell. This means that the ionic concentrations at the outer shell layer are larger than those at the inner surface. The reason for this is twofold. First, it will be entropically favorable for ions to stay outside the confining shell, secondly, the outer contact layer has a slightly larger radius of curvature than the contact surface at the inner side, and is thus able to adsorb a proportionally large number of ions (recall that there is an exclusion zone around the surface in which ions can not penetrate). Obviously, these effects will be more pronounced at larger salt concentrations. As the charge on the shell surface increases, the electrostatic stress starts to become important, leading to the swelling of capsid. Again, this effect is influenced by the amount of added electrolyte. At physiological salt concentrations ($\sim 150 $~mM) the compressed shell swells to its unstretched size ($u_r=0$) at about $Z\approx 200$, while in the case of large added salt concentration $c_s=800$~mM the relaxed state is only achieved only when $Z\approx 900$. 
	
	
\begin{figure}[h!]
		
		\centering
		\includegraphics[width=7.5cm,height=5.5cm]{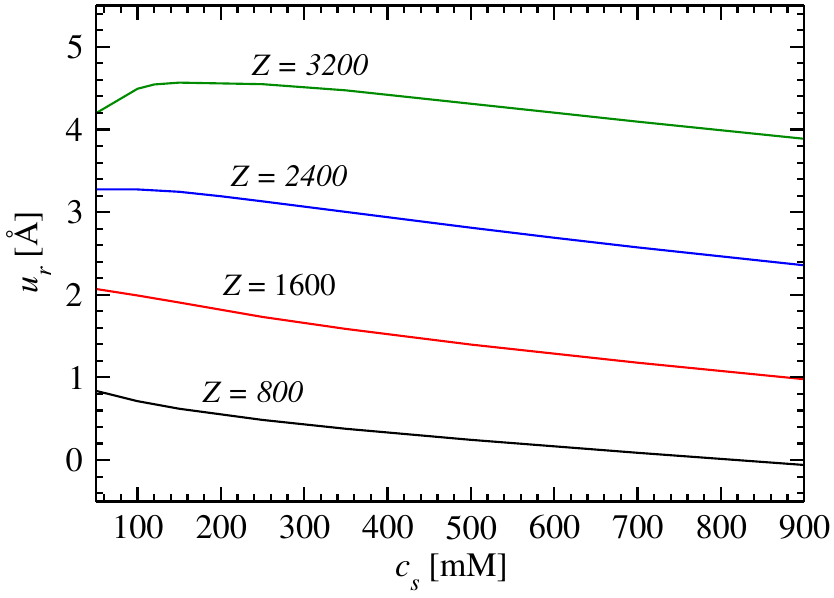}
		\caption{Radial strain $u_r=R-R_0$ as a function of the reservoir bulk concentration $c_s$ for different charges on the capsid surface. The equilibrium shell radius is $R_0=10$~nm, and the surface charges are displayed in the curves. }
		\label{fig:fig5}
	\end{figure}

	A deeper qualitative picture of the interplay between charge and salt concentration on the  shell mechanical equilibrium is provided by Fig (\ref{fig:fig5}). Here, radial strain as a function of the added salt concentration is displayed for different shell charges. Notice that in all cases the relative strain is less then $5\%$, even for large surface charge and small salt concentrations. The increase in charge always leads to the expansion of the shell. On the other hand the dependence on salt is not monotonic, see Fig (\ref{fig:fig5}). Addition of salt at large concentrations leads to shrinking (possible crumpling) of the shell. Remarkably, the relative swelling with the increase of salt concentration decreases linearly for salt concentrations above the physiological limit. 
	This trend is modified, however, for larger charges and small salt concentrations, when the strain initially grows upon increase of salt concentration. Again, this behavior can be attributed to a (weak) breakdown of charge neutrality ($Z_{in}\lesssim Z/2$) inside the shells with larger surface charges in solutions of low ionic strengths. For large surface charge, even a small deviation from charge neutrality can be enough to significantly change the interplay between the electrostatic and mechanical stresses, thereby leading to a shell inflation when salt is added and electroneutrality is reestablished. 
	
	
	\subsection{Pore Nucleation}
	
	In the analysis of the mechanical equilibrium performed so far, it was implicitly assumed that the shell will only change its size in response to the osmotic stress. This will be the case only if the shell is mechanically stable against the resulting strains. If not, the shell will not be able to withstand the shape deformation and will either disassemble or open a pore in order to release the elastic stress. If the degree of disassembly or the pore size becomes too large, the whole object might become structurally unstable and break apart or rupture. In order to address the issue of mechanical stability against the applied ionic osmotic stress, we here rely on a simple continuum approach similar in spirit to the Classical Nucleation Theory (CNT). Despite its simplicity, this  approach incorporates the main aspects that control the overall shell mechanical stability~\cite{Zan06}. 
	
	In order to open a pore or to disassemble building blocks from the surface, the bonds between the surface particles must be broken. If typical thermal fluctuations in the stretched surface are able to surmount such energetic barrier, the opening of a pore will take place spontaneously. This mechanism 
	resembles the nucleation of liquid drops in a vapor phase. It is very likely that the opening of such a hole in a surface (which, in the present case of a viral capsid corresponds to a removal of a cap) will be dynamically unstable, leading to a irreversible rupture of the whole structure~\cite{Idi04,Levin04}.
	
\begin{figure}[h!]
		\centering
		\includegraphics[width=9cm,height=6.5cm]{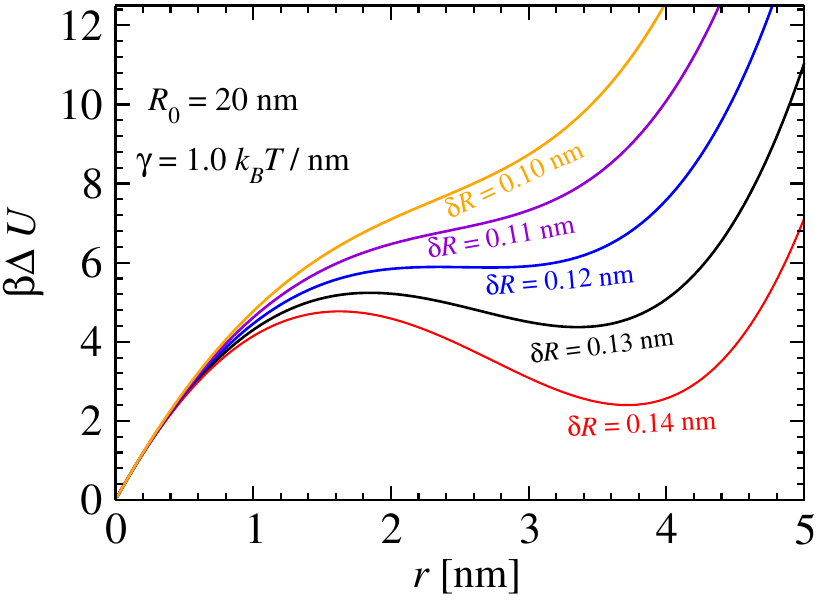}
		\caption{Elastic energy difference $\beta\Delta U=\beta U(r)-\beta U(0)$ for opening a pore of radius $r$. For large shell deformations $\delta R$, this function undergoes a maximum (activation barrier) before a minimum at pore sizes specified by Eqs. (\ref{rp1}) and (\ref{rp2}), respectively, is achieved. If the size deformations are smaller than a critical value ($\delta R_c \approx 0.12$~nm in this case, see blue line), the two local extrema disappear (after merging together at $\delta R_c$), and the shell becomes stable against pore nucleation. Here, the line tension is set to $\gamma=1~k_BT$/nm, while the unstressed shell size is $R_0=20$~nm. }
		\label{fig:fig6}
	\end{figure}

	In a continuum elastic approach, the opening of a circular pore of radius $r_p$ leads to a local release of the surface tension. In other words, the pore area $A_p=\pi r_p^2$ will becomes free of stretching forces \footnote{We suppose that the disassembled areas are small enough, such that curvature effects are negligible.}. The overall stretching energy $\beta U^{elas}$ thus becomes~\cite{Idi04}
	\begin{equation}
	\beta U^{elas}=\beta\dfrac{\kappa_s}{2A_0}(\delta A-A_p)^2,
	\label{U_elas2}
	\end{equation}  
	where $\delta A\equiv A-A_0$ is the surface area difference between stretched and unstretched shells. On the other hand, opening a hole in a surface obviously leads to a local breaking of bonds that originally keep the capsid particles together. In the framework of a mechanical continuum theory, the corresponding energy cost $U^{pore}$ is proportional to the size of the rim $2\pi r_p$ which becomes unbounded and exposed to the solution. The proportionality constant defines the line tension $\gamma$, which depends on specific material properties, such that $\beta U^{pore}=2\pi \beta \gamma r_p$. The total energy of a stretched shell with an open pore of radius $r_p$ on its surface is therefore $\beta U(r_p)=\beta U^{elas}=\beta\dfrac{\kappa_s}{2A_0}(\delta A-A_p)^2+2\pi \beta\gamma r_p$. To investigate the nucleation process, we must look at the energy difference between the membrane state with a pore and without \cite{Idi04}, 
	\begin{equation}
	\Delta \beta U \equiv \beta U(r_p)-\beta U(r_p=0)=\dfrac{\beta\kappa_s}{2A_0}\left(A_p^2-2A_p\delta A\right)+2\pi\beta\gamma r_p.
	\label{pore1}
	\end{equation} 
	The first term on the right-hand side is the surface energy difference between a fully closed shell and the one containing a hole of radius $r_p$.  The second term is the energy penalty for breaking the surface bonds and exposing the (usually hydrophobic) coating particles to the aqueous medium. Although this expression relies on a continuum approach, a straightforward one-to-one correspondence can be easily made with the case of spherical-like capsids with a relatively small number of building blocks. In that case, the first term can be interpreted as the energy difference between capsids in their fully and partially assembled states, whereas the second term represents the energy penalty resulting from the removal of capsomer neighboring bonds when a capsomer is released into the solution. The surface and pore areas can be easily related to the size of capsomers, the number of capsomers needed to assemble a closed capsid as well as the number of these entities in the partially formed capsid. The continuum quantities $\kappa_s$ and $\gamma$ are both in this case proportional to the binding energy that drives the capsid self-assembly.
	Eq. (\ref{pore1}) can be conveniently rewritten in terms of the dimensionless variables
	\begin{equation}
	\beta\Delta U =\dfrac{\tilde{\kappa}_s}{4}\left[\dfrac{\tilde{r}_p^4}{8}-\tilde{r}_p^2(\tilde{R}^2-1)\right]+2\pi\tilde{\gamma}\tilde{r}_p,
	\label{pore2}
	\end{equation} 
	where $\tilde{r}_p=r_p/R_0$ and $\tilde{R}=R/R_0$ are dimensionless pore and shell radii, respectively, $\tilde{\kappa}_s=\beta \kappa_sA_0$ is a dimensionless elastic modulus and $\tilde{\gamma}=\beta\gamma R_0$ a dimensionless line tension of the shell. For small pore sizes, the last term of this expression dominates, and the energy difference grows as $r_p$ increases~\cite{Idi04}. This means that microscopic pores are always energetically unfavorable.  
	Depending on the swelling ratio $\tilde{R}$, the term proportional to $\sim \tilde {r}_p^{2}$ might dominate at intermediate values of $\tilde{r}_p$, stabilizing larger pores, but requiring a thermal fluctuation to overcome the energy barrier in order to nucleate a pore~\cite{Idi04}, see Fig. \ref{fig:fig6}. The minimum of Eq.(\ref{pore2}) takes place at the most likely pore size to be nucleated once the energy barrier has been overcome. In the case of a viral capsid, the opening of such a cap in the surface will certainly lead to the rupture of the whole structure.  
	
	
	Viral capsids are generally robust against mechanical rupture. A clear evidence of such strong endurance is the large range in size polydispersity in which some virus capsids can be found. That clearly suggests that capsids should be able to undergo reasonably large deformations before disassembly takes place. In order to model the mechanical stability of these objects at different environmental conditions (i.e. surface charge and ionic strength), proper values should be assigned to the line tension $\gamma$. Notice that this quantity is proportional to the capsomer binding energy that promotes capsid self-assembly. This mean-field quantity should be high enough to ensure a reasonably degree of size polydispersity, and yet not too large so as to lead to dynamically-trapped assembling states during particle assembling. A rough estimation valid for a wide range of virus capsids stipulates that the binding energy per capsomer, $u_c$, lies within the range between $u_c\approx 10~k_BT$s and $u_c\approx 20~k_BT$s~\cite{Vij98,Zan05}.  For capsids with large number of capsomers, local curvature effects can be neglected, and a simple scaling argument can be employed to estimating a typical order of magnitude of $\gamma$. Assuming each capsid as a planar disc of radius $r_c$, the energy per length of binding contact is $\gamma=u_c/2\pi r_c$. For  a spherical capsid of radius $R_0$ and a total of $N$ capsomers, the radius $r_c$ of each capsomer can be approximate as $r\approx 2R_0/\sqrt{N}$ (assuming that the flat discs cover the whole surface area). This provides a rough estimate of $\gamma\approx u_c\sqrt{N}/(4\pi R_0)$ for the binding energy per length. Typically, the value of this parameter will, therefore, be smaller than one thermal energy per nanometer. For instance, a capsid with radius $R_0=25$~nm and a total of $N=180$ capsomers will have an estimated line tension of $\gamma\approx 0.64~k_BT$/nm.

\begin{figure}[h!]
		\centering
		\includegraphics[width=4.95cm,height=4.7cm]{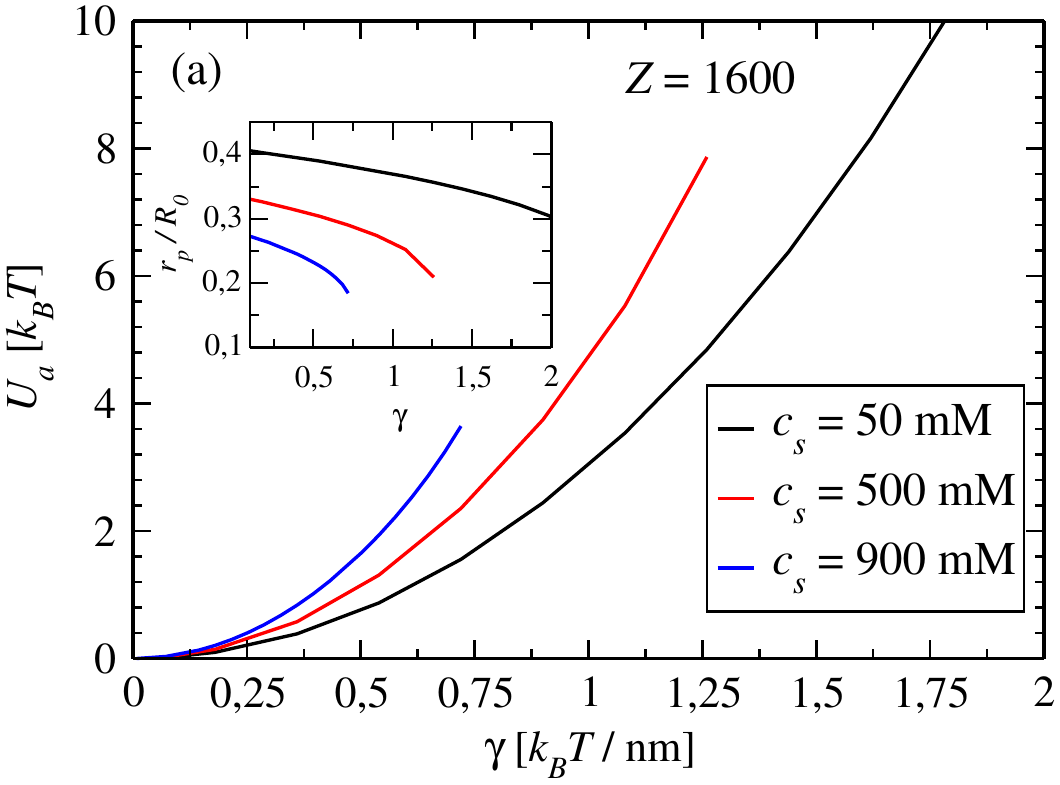}
		\includegraphics[width=4.95cm,height=4.7cm]{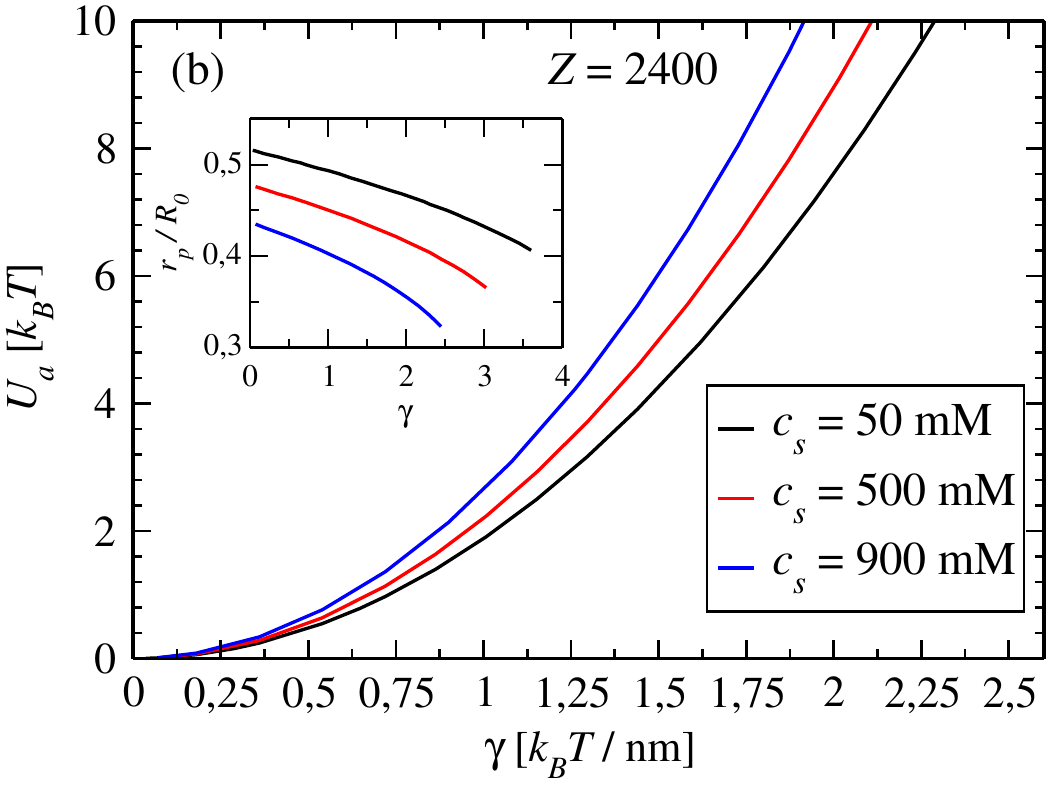}
		\includegraphics[width=4.95cm,height=4.7cm]{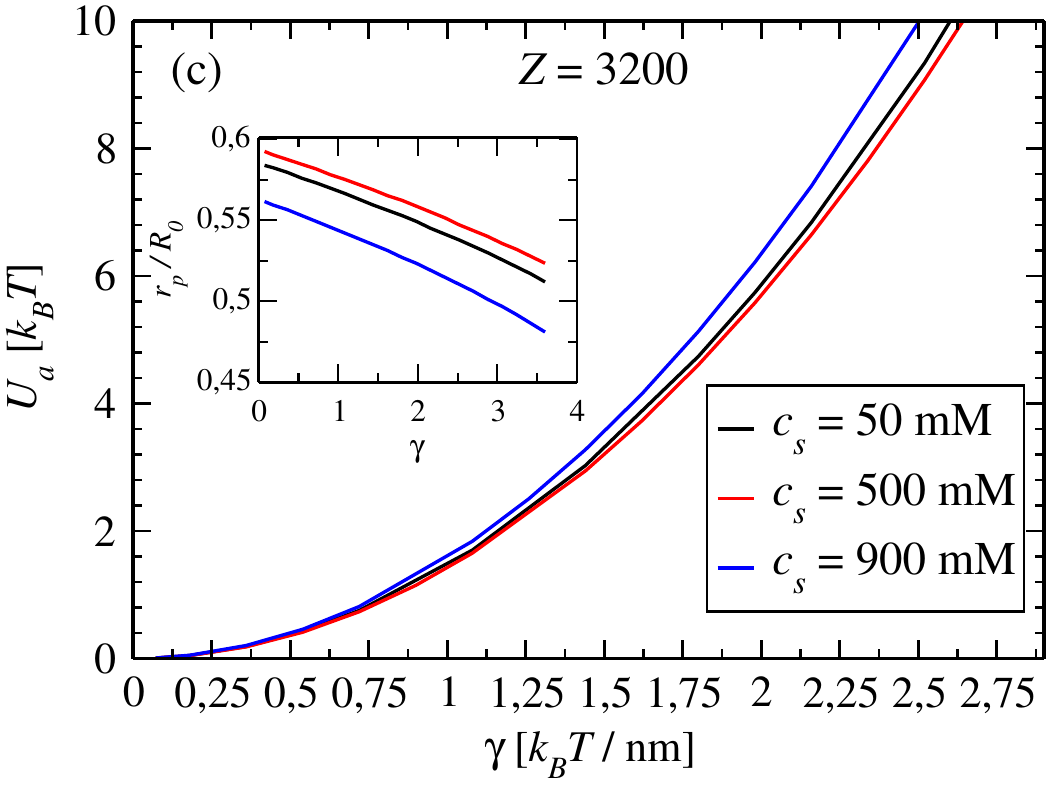}
		\caption{Activation energies $U_a$ for capsid disassembly as a function of the capsid line tension for different ionic concentrations. The capsid surface charges are:  $Z=1600$ (a); $Z=2400$ (b); and $Z=3200$ (c). The corresponding salt concentrations are indicated in the legends. The insets show the corresponding pore sizes when the pore nucleation takes place. The points where the lines end correspond to fully stable capsids without pores.}
		\label{fig:fig7}
	\end{figure}

	If the deformation $\delta A$ of the capsid is not too small, Eq. (\ref{pore2}) will display a maximum followed by a minimum, corresponding, respectively, to the activation barrier and a stable pore, as can be clearly identified in Fig.~\ref{fig:fig6}.  The overall behavior of the mechanical stability of viral capsids for a range of line tensions is summarized in Fig. \ref{fig:fig7}, where the activation energies (nucleation barriers) are shown for capsids bearing different charges and at different ionic strengths.  For small surface charges, the lines of nucleation barriers do not extend over all values of $\gamma$ (see Fig.~\ref{fig:fig7}a). This means that the minimum of Eq. (\ref{pore2}) disappears at these end-points, indicating that the capsids with larger values of $\gamma$ will be stable against disassembly. Notice that the end-points are shifted towards smaller values of $\gamma$ as the salt concentration increases. This implies that addition of salt will stabilize the capsids against rupture. Moreover, addition of salt leads to an increase of the activation barrier, rendering the system more robust against mechanical instability driven by the thermal fluctuations.

	
\begin{figure}[h!]
		\centering
		\includegraphics[width=7.4cm,height=5.5cm]{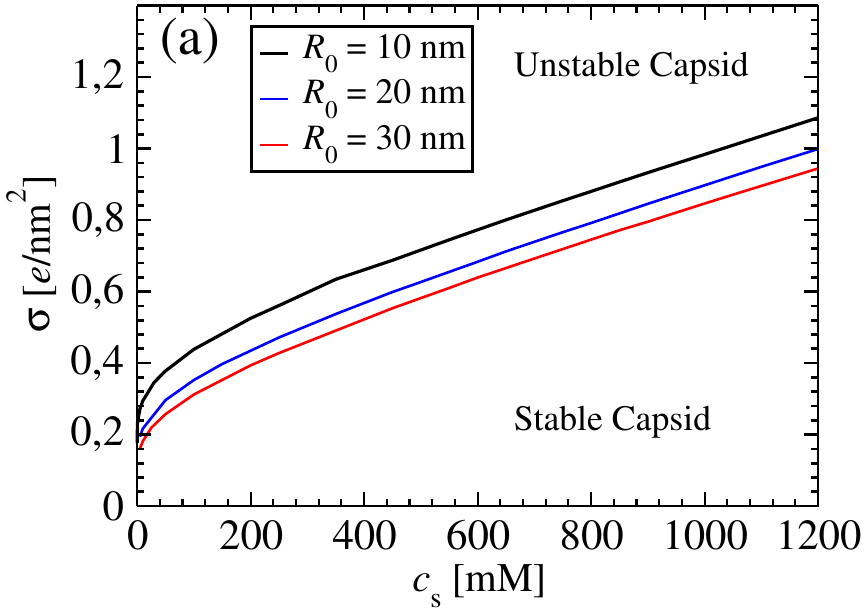}
		\includegraphics[width=7.4cm,height=5.5cm]{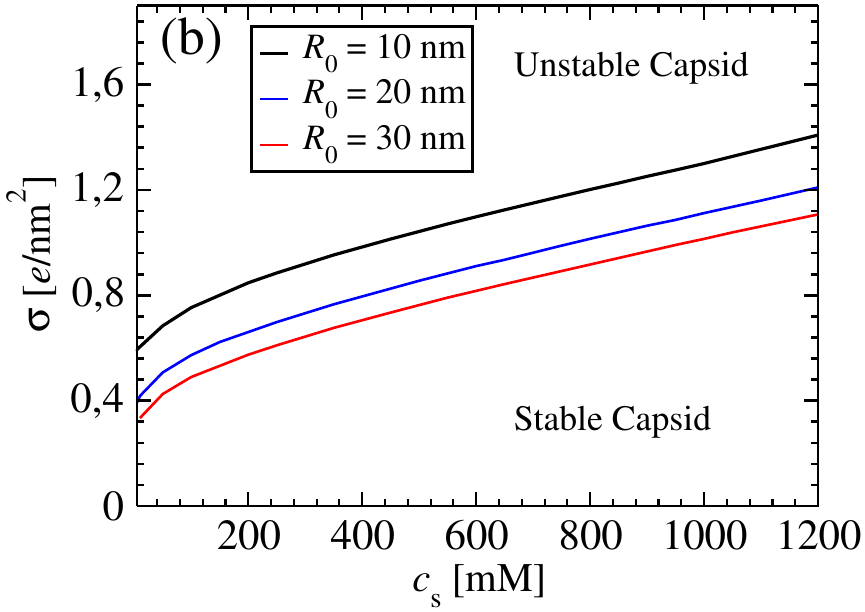}
		\caption{Stability phase diagram for shells of different size. The line tensions are: $\gamma=0.25~k_BT$/nm (a); and $\gamma=1.0~k_BT$/nm (b).}
		\label{fig:fig8}
	\end{figure}

	A close inspection of Eq. (\ref{pore2}) allows one to conclude that, at small relative deformations $(R-R_0)/R_0=\tilde{R}-1$, only a minimum at negative $r_p$ (therefore of no physical significance) occurs. For such parameters, no nucleation takes place, and capsids are mechanical stable against thermally-induced rupture~\cite{Idi04}. The parameter that controls the emergence of a nucleation barrier at small values of $\delta\tilde{R}-1$ is
	\begin{equation}
	\Delta\equiv\left(\dfrac{\tilde{R}^2-1}{3}\right)^3-\left(\dfrac{\pi\tilde{\gamma}}{\tilde{\kappa}}\right)^2.
	\label{Delta}
	\end{equation}  
	When $\Delta<0$, the energy difference in Eq. (\ref{pore2}) is a monotonically increasing function of $r_p$ for all $r_p>0$, and opening a pore is energetically unfavorable~\cite{Idi04}, see Fig. \ref{fig:fig6}. If $\Delta>0$, a maximum and a minimum of $\Delta U$ appear at \cite{Idi04}
	\begin{subequations}
		\begin{eqnarray}
		r_{p1} & = & 4\sqrt{\dfrac{\tilde{R}^2-1}{3}}\cos\left(\dfrac{\theta_0-2\pi}{3}\right)\label{rp1}\\
		r_{p2} & = & 4\sqrt{\dfrac{\tilde{R}^2-1}{3}}\cos\left(\dfrac{\theta_0}{3}\right)\label{rp2}
		\end{eqnarray}
	\end{subequations}
	respectively, where $\theta_0=\cos^{-1}\left(\dfrac{-2\pi\tilde{\gamma}}{\tilde{\kappa}(\tilde{R}^2-1)^{3/2}}\right)$, and the capsid will be mechanically unstable in this region (which corresponds to the stable curves above the blue line in Fig.~\ref{fig:fig6}). From this analysis, we can conclude that the end-points in Fig.~\ref{fig:fig7} that delimit mechanical stability will be such that 
	\begin{equation}
	\tilde{R}=\sqrt{1+3\left(\dfrac{\pi\tilde{\gamma}}{\tilde{\kappa}}\right)^{2/3}}.
	\label{Delta}
	\end{equation}  
	For fixed values of $\tilde{\kappa}$ and $\tilde{\gamma}$, this expression provides the threshold swelling size $\tilde{R}$ at which the capsid becomes mechanically unstable. 
	The critical deformation ratio $\tilde{R}$ can be inserted into Eq. (\ref{Pi_osm}), which then can be solved using the DFT approach discussed above to obtain points $(c_s,Z)$ that delimit the region of the mechanical stability of capsids. We have done this for two representative values of the line tension, $\gamma=0.25~k_BT$/nm and $\gamma=1.0~k_BT$/nm, considering three distinct capsid sizes of $R_0=10$~nm, $R_0=20$~nm, and $R_0=30$~nm, which can model a broad class of viral capsids. The DFT approach allows us to span a wide region in the $(c_s,Z)$ plane, yet keeping a high degree of accuracy, far beyond the range of validity of traditional mean-field theories. The results for the transition lines are shown in Fig.~\ref{fig:fig8}.  The lines corresponding to different shell sizes behave remarkably similarly, specially in the situation of small line tension $\gamma=0.25~k_BT$/nm (Fig. \ref{fig:fig8}a), a narrow region can be identified which delimits stable and unstable shells of various sizes. In the case of high line tension of $\gamma=1.0~k_BT$/nm (Fig. \ref{fig:fig8}b), the region between different lines becomes broader. As expected, an increase in the shell charge density requires a larger salt concentration in order to keep the capsid integrity. This is clearly due to the competing effects from various contributions: increasing the surface charge leads to a larger electrostatic repulsion between the capsomers, while increasing salt concentration results in a stronger ionic condensation at the outer layer, providing a compression force on the shell. Apart from  small salt concentrations, this behavior is linear in character. Moreover, the lines corresponding to different shell sizes are almost parallel to one another, suggesting a universal linear behavior with a slope that scales with the line tension.  
	
	
	\section{Conclusions}
	
	We have proposed a model which allows us to investigate the mechanical equilibrium and stability of biological nanoshells of spherical shape under a wide range of surface charges and salt concentrations.  The ionic equilibrium properties have been obtained using a DFT that accurately incorporates size and electrostatic effects, both important to properly capture the fine details of the inner and outer electric double layers that build-up across a charged shell. A simple analytical expression was then derived which relates the osmotic stress -- comprising ionic and electrostatic self-energy contributions -- in terms of quantities readily accessible in the underlying DFT approach. By combining this osmotic stress with a stretching Hook-like surface stress evaluated in the framework of a continuum elastic model, a mechanical equilibrium condition was numerically solved to obtain the equilibrium radial strains at different salt concentrations and shell charges. It was shown that at large surface charges electrostatic contributions dominate, whereas at moderate charges and salt concentrations a shell becomes compressed due to counterions condensed at the outer shell surface.  
	
	In order to address the important question of whether the obtained mechanical equilibrium states are stable against pore opening or coat particle disassembly, the mechanical approach is further combined with a simple elastic theory for pore nucleation. Opening of a pore on the shell surface takes place at a cost of line energy that accounts for breaking-up of surface bonds. This manifests itself as a nucleation barrier that, once surmounted {\it via} thermal fluctuations, can lead to an irreversible rupture of the shell. The approach allows us to identify the stable capsids.  The transition lines that separate stable and unstable shells are clearly identified for different shell sizes. These lines display a very similar linear behavior at large salt concentrations for all observed capsid sizes, indicating a universal behavior of different capsids possessing similar binding energies. 
	The model  can be further improved to extend its range of applicability. For example, the simplified approximation of a structureless thin shell can be easily replaced by a more realistic model of a membrane of finite  thickness. In this case, the mechanical bending rigidity should also be incorporated into the mean-field elastic description. Moreover, since the proteins on both layers of the capsid typically bear different charges, a more detailed picture would require shells with heterogeneous charge distribution on internal and external surfaces. Another important issue to address is the stability of capsids bearing a charged cargo -- representing either synthetic nanomaterials or a packaged DNA or RNA. In the latter case, an additional contribution from the bending energy of the compressed chains has to be somehow incorporated into the present model system~\cite{Lev02,Tri16}. Again, such an extension can be readily incorporated within the developed formalism and will be a subject of future work.

	\section{Acknowledgments}
	
	This work was partially supported by CNPq, CAPES, INCT-FCx, and  the US-AFOSR under the grant FA9550-12-1-0438.

	\newpage
	
	\section{Supplementary Information}

	\subsection{FMT in spherical geometry}
	
	We now provide explicit formulas for the FMT weighted densities in the underlying spherical geometry. The scalar weighted densities used in the hard-sphere functional can be written as
	\begin{equation}
	n_{\alpha}(\br)=\sum_i\int\rho_i(\br')\omega^{(i)}_{\alpha}(\br-\br')d\br'.
	\label{wd_1}
	\end{equation}  
	Here, the subscript $\alpha$ denotes the set of weighted function, whereas the upper index $(i)$ refers to the ionic spices. The weight functions can be either scalar or vector entities. The scalar ones are given by
	\begin{subequations}
		\begin{align}
		\omega_{3}^{(i)}(\br) & =  \Theta(a_i-r) \label{w1a}\\
		\omega_{2}^{(i)}(\br) & =  \delta(r-a_i)  \label{w1b}\\
		\omega_{1}^{(i)}(\br) & =  \dfrac{\omega^{(i)}_2(r)}{4\pi a_i}  \label{w1c}\\
		\omega_{0}^{(i)}(\br) & =  \dfrac{\omega^{(i)}_2(r)}{4\pi a_i^2}. \label{w1d}
		\end{align}
	\end{subequations}
	The subscript $\alpha$ is such that ($\alpha-3$) refers to the spatial dimensionality of the underlying weighted density. The vector weighted densities are represented as convolutions similar to the ones in Eq. (\ref{wd_1}), with the weight functions replaced by the following vectors:
	\begin{subequations}
		\begin{align}
		\bom_{2}^{(i)}(\br) & =  -\nabla\omega_{3}^{(i)}(\br)=\delta(r-a_i)\be_r \label{w2a}\\
		\bom_{1}^{(i)}(\br) & =  \dfrac{\bom_{2}^{(i)}(\br)}{4\pi a_i},\label{w2b}
		\end{align}
	\end{subequations}
	where $\be_r=\br/r$ is the unit vector pointing at the radial direction. In the present situation of radially symmetric potentials, the density profiles depend only on the radial coordinate $r'$. Likewise, the scalar weight functions depend only on the relative distance $R=|\br-\br'|=\sqrt{r^2+r'^2-2\br\cdot\br'}$ between source and observation points. The integrals in (\ref{wd_1}) can therefore be explicitly written as
	\begin{equation}
	n_{\alpha}(\br)=\sum_i\int_0^{\infty}\rho_i(r')r'^2dr'\int_0^{2\pi}d\varphi'\int_0^{\pi}\omega^{(i)}_{\alpha}(R)\sin\theta' d\theta'.
	\label{wd_2}
	\end{equation}  
	
	If we now conveniently set the $z$-axis along the direction of the observation point $\br$ in performing the above integral, the relative distance $R$ becomes $R=\sqrt{r^2+r'^2-2rr'\cos\theta'}$. Integration over the azimuthal angle $\varphi'$ can be readily performed, while the integration over polar angle $\theta'$ can be converted into an integral over the relative distance $R$. To this end, we note that $RdR=\dfrac{\sin\theta'}{rr'}d\theta'$. Converting the integration limits accordingly, we arrive at the following result for the scalar weighted densities:
	\begin{equation}
	n_{\alpha}(r)=\dfrac{2\pi}{r}\sum_i\int_{0}^{\infty}r'\rho_i(r')dr'\int_{|r-r'|}^{r+r'}\omega^{(\alpha)}_i(R)RdR.
	\label{radial1}
	\end{equation}
	A similar reasoning can be applied to rewrite the vector weight densities
	\begin{equation}
	\bm{n}_{\alpha}(r)=\sum_i\int\bom_{\alpha}^{(i)}(\br-\br')\rho_i(\br')d\br'
	\label{wd_3}
	\end{equation}
	into a much simplified form. We first notice that the vector weight functions in Eqs. (\ref{w2a}) and (\ref{w2b}) can be written as $\bom_{\alpha}^{(i)}(\br-\br')=|\bom_{\alpha}^{(i)}(R)|\be_R$, where $\be_R=\frac{(\br-\br')}{R}$ is the unit vector connecting integration and observation points. The integrals above can thus be explicitly written as
	\begin{equation}
	\bm{n}_{\alpha}(r)=\sum_i\int_0^{\infty}\rho_i(r')r'^2dr'\int_0^{\pi}\sin\theta' d\theta'\int_0^{2\pi}\dfrac{|\bom_{\alpha}^{(i)}(R)|}{R}(\br-\br')d\varphi'.
	\label{wd_3}
	\end{equation}
	Once again, it is convenient set the (fixed) radial vector $\be_r$ as pointing along the $z$-axis, $\be_z$, while performing the above integral. With this choice, the integration source point $\br'$ can be composed in terms of its cartesian components as 
	\begin{equation}
	\br'=r'(\sin\theta'\cos\varphi'~\be_x+\sin\theta'\sin\varphi'~\be_y+\cos\theta'~\be_z),
	\label{source}
	\end{equation}
	while $\br=r\be_r=r\be_z$. It is easy to check that contributions in the $x$ and $y$ directions will vanish when the above expression is inserted into Eq. (\ref{wd_3}), since the azimuthal integrals are zero. Only the contribution along the $z$-axis (which coincides with the $\be_r$ direction) survives, and a simple integration over the azimuthal angle provides
	\begin{equation}
	\bm{n}_{\alpha}(r)=2\pi\be_r\sum_i\int_0^{\infty}\rho_i(r')r'^2dr'\int_0^{\pi}\dfrac{|\bom_{\alpha}^{(i)}(R)|}{R}\left(r-r'\cos\theta'\right)\sin\theta' d\theta'.
	\label{wd_4}
	\end{equation}
	The second integral over polar angle can be again transformed into an integral over the relative distance $R$, under the simple replacement $\cos\theta'=\dfrac{r^2+r'^2-R^2}{2rr'}$. The above integral assumes then the form
	\begin{equation}
	\bm{n}_{\alpha}(r)=\dfrac{\pi}{r^2}\be_r\sum_i\int_{0}^{\infty}r'\rho_i(r')dr'\int_{|r-r'|}^{r+r'}|\bom_{\alpha}^{(i)}(R)|\left[R^2+r^2-r'^2\right]dR. 
	\label{radial2}
	\end{equation}
	
	It is quite clear from Eqs. (\ref{radial1}) and (\ref{radial2}) that both scalar and vector weighted densities will be radially symmetric, just like the original densities. Besides, the vector densities always point in the radial direction of the observation point, $\be_r$. Inserting the weight functions from Eqs. (\ref{w1b}) and (\ref{w2a}) into Eqs. (\ref{radial1}) and (\ref{radial2}), respectively, leads to the following explicit relations:
	\begin{subequations}
		\begin{eqnarray}
		n_{2}(r) & = & \dfrac{2\pi}{r}\sum_i \int_{0}^{\infty}r'\rho_i(r')dr'\int_{|r-r'|}^{r+r'}\delta(R-a_i)RdR\label{bn2a}\\
		\bn_{2}(r) &  = &  \dfrac{\pi}{r^2}\be_r\sum_i \int_{0}^{\infty}r'\rho_i(r')dr'\int_{|r-r'|}^{r+r'}\delta(R-a_i)\left(R^2+r^2-r'^2\right)dR\label{bn2b}.
		\end{eqnarray}
	\end{subequations}
	The integrals over the relative distance $R$ will clearly vanish whenever the point $R=a_i$ lies outside the range of integration. If $r>a_i$, this condition is fulfilled for $r'$ in the range $r-a_i\le r'\le r+a_i$ (for the upper integration limit is always bigger than $a_i$ in this case). On the other hand, if $r<a_i$, this condition implies $a_i-r\le r'\le r+a_i$. Thus, only values of $r'$ within these ranges will have a non-vanishing contribution in the first integrals above. Moreover, the delta functions will simply filter the points $R=a_i$ in these intervals, resulting in the following simplified expressions:
	\begin{subequations}
		\begin{eqnarray}
		n_{2}(r) & = & \dfrac{2\pi}{r}\sum_i a_i\int_{|r-a_i|}^{(r+a_i)}r'\rho_i(r')dr'\label{bn2a}\\
		\bn_{2}(r) &  = & \dfrac{\pi}{r^2}\be_r\sum_i\int_{|r-a_i|}^{(r+a_i)}r'\rho_i(r')[r^2+a_i^2-r'^2]dr'\label{bn2b}.
		\end{eqnarray}
	\end{subequations}
	There is an apparent singularity in the above weighted functions as one approaches the center of the shell ({\it i. e.}, at $r\rightarrow 0$). However, it is easy to check that the integrals in (\ref{bn2a}) and (\ref{bn2b}) scale as $\sim r$ and $\sim r^3$, respectively, at this point, so that the weighted densities remain finite at the origin. From the above expressions, explicit relations for the weighted densities $n_0(r)$, $n_1(r)$, as well as for the vector density $\bn_{1}$ follow direct by using Eqs. (\ref{w1d}), (\ref{w1c}) and (\ref{w2b}), respectively. The results are:
	\begin{subequations}
		\begin{eqnarray}
		n_{0}(r) & = & \dfrac{1}{2r}\sum_i \dfrac{1}{a_i}\int_{|r-a_i|}^{(r+a_i)}r'\rho_i(r')dr'\label{bn3a}\\\nonumber\\
		n_{1}(r) & = & \dfrac{1}{2r}\sum_i\int_{|r-a_i|}^{(r+a_i)}r'\rho_i(r')dr'\label{bn3b}\\\nonumber\\
		\bn_{1}(r) & = & \dfrac{1}{4r^2}\be_r\sum_i\dfrac{1}{a_i}\int_{|r-a_i|}^{(r+a_i)}r'\rho_i(r')[r^2+a_i^2-r'^2]dr'\label{bn3c}.
		\end{eqnarray}
	\end{subequations}
	Now, the remaining weighted density $n_{3}(\br)$ can be obtained by inserting the weight function (\ref{w1a}) into (\ref{radial1}). Explicitly, one gets:
	\begin{equation}
	n_{3}(r)=\dfrac{2\pi}{r}\sum_i\int_0^{\infty}r'\rho_i(r')dr'\int_{|r-r'|}^{r+r'}\Theta(R-a_i)RdR.
	\label{n3_1}
	\end{equation}
	Notice that the last integral vanishes in the region $|r-r'|>a_i$. When $r>a_i$, this implies that the only non-vanishing contributions come from $r'$ in the region $r-a_i<r'<r+a_i$ (note that the upper integration limit is always greater than $a_i$ in this case), whereas if $r\le a_i$ the non-vanishing contributions come from $a_i-r\le r'<r+a_i$.  Moreover, if $r+r'>a_i$, this upper integration limit is to be replaced by $a_i$. Clearly, this will always happen in region $r>a_i$. Combining these results, we can split the above integral into such distinct regions as follows:
	\begin{equation} 
	n_{3}(r)  = \begin{cases}
	\displaystyle{\dfrac{2\pi}{r}\sum_i\left[\int_{0}^{a_i-r}r'\rho_i(r')dr'\int_{|r-r'|}^{r+r'}RdR+\int_{a_i-r}^{r+a_i} r'\rho_i(r')\int_{|r-r'|}^{a_i}RdR\right],\hspace{1cm}r\le a_i,}\\ \\
	\displaystyle{\dfrac{2\pi}{r}\sum_i\int_{r-a_i}^{r+a_i} r'\rho_i(r')dr'\int_{|r-r'|}^{a_i}RdR,\hspace{5.7cm}r\ge a_i.}
	\end{cases} 
	\label{n3_2}
	\end{equation}
	Now, the integrals over $R$ can be readily performed, and the above expressions finally simplify to:
	\begin{equation} 
	n_{3}(r)  = \begin{cases}
	\displaystyle{\dfrac{\pi}{r}\sum_i\left[4r\int_{0}^{a_i-r}r'^2\rho_i(r')dr'+\int_{a_i-r}^{r+a_i} r'\rho_i(r')[a_i^2-(r-r')^2]dr'\right],\hspace{1cm}r\le a_i,}\\ \\
	\displaystyle{\dfrac{\pi}{r}\sum_i\int_{r-a_i}^{r+a_i} r'\rho_i(r')[a_i^2-(r-r')^2]dr',\hspace{5.7cm}r\ge a_i.}
	\end{cases} 
	\label{n3_2}
	\end{equation}
	
	Again, it is important to note that this weight function remains finite at the origin, since the second integral in the first line above has leading term proportional to $\sim r$ in this limit. Notice also that by virtue of the identity in (\ref{w2a}), Eq. (\ref{bn2b}) can be obtained from the above equation by making $\bn_{3}(r)=-\nabla n_{3}(r)$. 
	
	The expressions provided above show that numerical integration to obtain the weighted densities can be effectively performed considering only one-dimensional integrals over a small region of at most one diameter size around each observation point $r$. After numerical calculation of the weighted densities, the hard-sphere interaction contribution to the excess chemical potential can be readily computed using:
	
	\begin{equation}
	\beta\mu_i(\br)=\dfrac{\delta\beta\mathcal{F}^{hc}}{\delta\rho_i(\br)}=\sum_{\alpha}\int\mu_{\alpha}(\br')\dfrac{\delta n_{\alpha}(\br') }{\delta\rho_i(\br)}d\br',
	\end{equation}
	where we have defined $\mu_{\alpha}(\br)\equiv\dfrac{\partial \Phi}{\partial n_{\alpha}}\biggr\arrowvert_{n_{\alpha}(\br)}$ as the derivative of the (local) free-energy density in the FMT functional with respect to the weighted ionic densities. Using Eq. (\ref{wd_1}), the expression above can be simplified to
	
	\begin{equation}
	\beta\mu_i(\br)=\sum_{\alpha}\int\beta\mu_{\alpha}(\br')\omega^{(i)}_{\alpha}(\br'-\br)d\br'.
	\label{mu_hs}
	\end{equation}
	
	In the case of vector weight functions, the above integrals are generalized to a scalar product between the gradient of $\Phi(n_{\alpha})$ with respect to the components of the vector density $\bm{n}_{\alpha}$ and the corresponding weight density $\bom_{\alpha}$. Since the vectors $\bm{\mu}_{\alpha}(\br')$ point in the radial direction $\bm{\hat{e}}_{r'}$, whereas the weight densities point along the direction of $-\be_R=(\br'-\br)/R$, these integrals can be written as
	\begin{equation}
	\int\bm{\mu}_{\alpha}(\br')\cdot\bom_{\alpha}^{(i)}(\br'-\br)d\br'=\int\dfrac{|\bm{\mu}_{\alpha}(\br')|}{R}|\bom_{\alpha}^{(i)}(R)|\left(r'-r\cos\theta'\right)d\br',
	\label{mu_hs2}
	\end{equation}
	where again $\theta'=\cos^{-1}(\bm{\hat{e}}_{r}\cdot\bm{\hat{e}}_{r'})$ is the angle between the vectors $\br$ and $\br'$. As before, we can set the $z$-axis so as to coincide with the observation point direction $\br$. The azimuthal integration can thus be trivially performed, while the integration of polar angle can be simplified under the substitution $\cos\theta'=(r^2+r'^2-R^2)/(2rr')$. The above expressions are then simplified to
	\begin{equation}
	\int\bm{\mu}_{\alpha}(\br')\cdot\bom_{\alpha}^{(i)}(\br'-\br)d\br'=\dfrac{\pi}{r^2}\int_0^{\infty}|\bm{\mu}_{\alpha}(r')|r'dr'\int_{|r-r'|}^{r+r'}|\bom_{\alpha}^{(i)}(R)|\left(r^2+R^2-r'^2\right)dR.
	\label{mu_hs22}
	\end{equation}
	
	Note that, because the weighted densities $n_{\alpha}(r)$ all possess radial symmetry, the functions $\mu_{\alpha}$ will be also spherical symmetric, as well as the resulting chemical potentials in Eq. (\ref{mu_hs}). As a consequence, all the integrals in each term of this expression can be simplified to a one-dimensional radial integral, $I_{\alpha}^{(i)}(r)$, whose form is identical to the corresponding $n_{\alpha}(r)$ integrals given above, provided the simple replacement $\rho_{i}(r)\leftrightarrow\mu_{\alpha}(r)$ is made. 
	
	As a final remark we notice that, since the numerical integrals are in practice performed over a finite volume, the upper integration limits over the radial coordinate $r'$ are to be replaced by $R_{min}=min(R_c,r+a_i)$, where $R_c$ is the radius of the confining cell in which integration is performed. 
	
	\section{Force balance across the shell}

	We now provide a detailed derivation of the force-balance condition across the spherical charged shell of radius $R$. Since the system possess  spherical symmetry, the net force on an arbitrary point on the shell surface will point in the radial direction. This force can be either positive or negative, resulting in an outward or inward
	osmotic stress, respectively. The net force on the shell is the force exerted by the surrounding ionic cloud on its surface. On the other hand, the force due to the electrolyte on the shell is the negative of the force that the shell exerts on the ionic system. Due to the spherical symmetry, the force $dF$ acting on each element of area $dA$ on the shell surface is the same. The corresponding pressure is therefore $P=\dfrac{dF}{dA}$. On the other hand, the net force on the wall can be split into electrostatic and hard-sphere contributions. Making use of the spherical symmetry, the electrostatic contribution to the osmotic stress over the surface is
	\begin{equation}
	{\Pi}_s^{el}=\dfrac{1}{A}q\int\varrho_s(\br)(\bm{E}_{ion}(\br)\cdot{\be_r})d\br,
	\label{Fs1}
	\end{equation}
	where $\varrho_s(r)=Zq\delta(r-R)/4\pi R^2$ is the charge density lying on the shell surface, $\bm{E}_{ion}$ is the electric field produced by the mobile ions only, $\be_r$ is the unit vector pointing at the radial direction and $A=4\pi R^2$ is the surface area. Notice that, while the {\it net} force on the shell is obviously zero, the radial force on an arbitrary point on the surface does not vanish. Since the ionic profiles have radial symmetry, application of the Gauss Law allows one to write the ionic electric field as $\bm{E}_{ion}(r)=Z_{ion}(r)q/\varepsilon r^2\be_r$, where $Z_{ion}(r)$ is the total ionic charge enclosed within a sphere of radius $r$. Substituting  these results in the above expression provides the following expression for the electrostatic pressure on a given point on the surface:
	\begin{equation}
	\Pi_{s}^{el}=\dfrac{qZE_{ion}(R)}{A}=\dfrac{ZZ_{in}\lambda_B}{4\pi R^4},
	\label{Pi_el}
	\end{equation}
	where we have defined $Z_{in}\equiv Z_{ion}(R)$ as the net ionic charge lying {\it inside} the spherical shell. Since the net ionic charge inside the shell volume has sign opposite to the shell surface charge surface, this contribution to the osmotic stress is usually negative, leading to the shrinkage of the shell surface.
	
	Let us now consider the hard-core ion-wall interaction to the osmotic pressure. According to Newton's third Law, the net radial pressure due to ionic collisions at close contact with the shell membrane can be expressed as
	\begin{equation}
	\Pi_s^{hs}=\dfrac{1}{A}\sum_i\int\rho_i(\br)\left(\nabla\phi^{hs}_i(\br)\cdot\be_r\right)d\br, 
	\label{F_hs1}
	\end{equation}
	where $\phi^{hs}_i(\br)$ is the ion-shell hard-core potential. Using the radial symmetry of both ionic profiles and ion-shell hard-core interactions, the expression above can be conveniently rewritten as
	\begin{equation}
	\beta{\Pi}_s^{hs}=-\dfrac{4\pi}{A}\sum_i\int_0^{\infty}r^2\dfrac{d}{dr}\left(e^{-\beta\phi_i^{hs}(r)}\right)\rho_i(r)e^{\beta\phi_i^{hs}(r)}dr.
	\label{radF8}
	\end{equation}
	Note that, in contrast to $\phi_i^{hs}(r)$, the quantity $e^{-\phi^{hs}_i}$ is limited everywhere. This function vanishes at ion-shell overlap, being equal to unity anywhere else. Integration by parts of the above expression yields
	
	\begin{equation}
	\beta{Pi}^{hs}=-\dfrac{4\pi}{A}\sum_i\left[\int_0^{\infty}\dfrac{d}{dr}\left(\rho_i(r)r^2\right)dr-\int_0^{\infty}e^{-\beta\phi_i^{hs}(r)}\dfrac{d}{dr}\left(\rho_i(r)r^2e^{\beta\phi_i^{hs}(r)}\right)dr\right].
	\label{F_hs2}
	\end{equation}
	Now, noticing that the quantity $\phi_i^{hs}(r)$ vanishes in the regions of non-overlapping, it becomes clear that the integrals above cancel each other in such regions (since $e^{\beta\phi_i^{hs}}=e^{-\beta\phi_i^{hs}}=1$ there). Moreover, the second integral vanishes when ion and shell overlap. The only contribution left is, therefore,
	\begin{equation}
	\beta{Pi}^{hs}=-\dfrac{4\pi}{A}\sum_i\int_{R_-}^{R_+}\dfrac{d}{dr}\left(\rho_i(r)r^2\right)dr=\dfrac{4\pi}{A}\sum_i\rho_i(R_{-})R_{-}^2-\rho_i(R_{+})R_{+}^2,
	\label{F_hs3}
	\end{equation}
	where $R_{\pm}$ denotes the closest inner/outer ion-shell contact distance. For very thin shells $R_{\pm} \approx R$. The radial contribution from ion-shell hard-core interactions to the osmotic pressure is then 
	\begin{equation}
	\beta\Pi^{hs}=\sum_i\rho_i(R_{-})-\rho_{i}(R_+).
	\label{Pi_hs}
	\end{equation}
	Note that the inner (outer) contact ionic densities dictate the outward (inward) contributions to the osmotic pressure. The overall ionic contribution to the osmotic stress can be obtained by combining of electrostatic and the hard-sphere contributions, Eqs. (\ref{Pi_el}) and (\ref{Pi_hs}), respectively, 
	\begin{equation}
	\beta\Pi^{osm}=\sum_i\rho_i(R_{-})-\rho_i(R_{+})+\dfrac{ZZ_{in}\lambda_B}{4\pi R^4}.
	\label{Pi_osm1}
	\end{equation} 
	The above expression comprises only {\it ionic} contributions to the osmotic stress, resulting from  corresponding the ion-shell interactions. The {\it total} osmotic stress should also contain the contribution from the shell electrostatic and elastic self-energies. The electrostatic self-energy is
	\begin{equation}
	\beta U_s^{self}=\dfrac{\varepsilon}{8\pi}\int|\bm{E}_s(\br)|^2d\br,
	\label{Us}
	\end{equation} 
	where $\bm{E}_s$ stands for the electric field produced by the charged shell. This field vanishes inside the charged shell, while at distances larger than the shell radius it is given by $\bm{E}_s(\br)=Zq/\varepsilon r^2\be_r$. Substitution of this expression into the above integral results in  $\beta U_s^{self}=\lambda_B Z^2/2R$. The corresponding contribution to the osmotic stress can be computed from $\beta\Pi=-\dfrac{1}{4\pi R^2}\dfrac{\partial\beta U_s^{self}}{\partial R}$, resulting
	in 
	\begin{equation}
	\beta\Pi^{self}_s=\dfrac{\lambda_BZ^2}{8\pi R}.
	\label{Pi_self}
	\end{equation}
	Note that this contribution is always positive. The total electrostatic and hard-sphere contribution to the osmotic stress can finally be written as
	\begin{equation}
	\beta\Pi_s=\sum_i\rho_{i}(R_{-})-\rho_{i}(R_{+})+\lambda_B\dfrac{Z(Z+2Z_{in})}{8\pi R^4}.
	\label{Pi_self}
	\end{equation}


\begin{thebibliography}{112}
		\expandafter\ifx\csname natexlab\endcsname\relax\def\natexlab#1{#1}\fi
		\expandafter\ifx\csname bibnamefont\endcsname\relax
		\def\bibnamefont#1{#1}\fi
		\expandafter\ifx\csname bibfnamefont\endcsname\relax
		\def\bibfnamefont#1{#1}\fi
		\expandafter\ifx\csname citenamefont\endcsname\relax
		\def\citenamefont#1{#1}\fi
		\expandafter\ifx\csname url\endcsname\relax
		\def\url#1{\texttt{#1}}\fi
		\expandafter\ifx\csname urlprefix\endcsname\relax\def\urlprefix{URL }\fi
		\providecommand{\bibinfo}[2]{#2}
		\providecommand{\eprint}[2][]{\url{#2}}
		
		\bibitem[{\citenamefont{de~la Escosura-Mu\~niz and Merko\c{c}i}(2012)}]{Mun12}
		\bibinfo{author}{\bibfnamefont{A.}~\bibnamefont{de~la Escosura-Mu\~niz}}
		\bibnamefont{and}
		\bibinfo{author}{\bibfnamefont{A.}~\bibnamefont{Merko\c{c}i}},
		\bibinfo{journal}{ACS Nano} \textbf{\bibinfo{volume}{6}},
		\bibinfo{pages}{7556} (\bibinfo{year}{2012}).
		
		\bibitem[{\citenamefont{Karp}(1999)}]{karp}
		\bibinfo{author}{\bibfnamefont{G.}~\bibnamefont{Karp}},
		\emph{\bibinfo{title}{Cell and molecular biology: concepts and experiments}}
		(\bibinfo{publisher}{New York: J. Wiley}, \bibinfo{year}{1999}).
		
		\bibitem[{\citenamefont{Deserno et~al.}(2013)\citenamefont{Deserno, Kremer,
				Paulsen, Peter, and Schmid}}]{Deserno}
		\bibinfo{author}{\bibfnamefont{M.}~\bibnamefont{Deserno}},
		\bibinfo{author}{\bibfnamefont{K.}~\bibnamefont{Kremer}},
		\bibinfo{author}{\bibfnamefont{H.}~\bibnamefont{Paulsen}},
		\bibinfo{author}{\bibfnamefont{C.}~\bibnamefont{Peter}}, \bibnamefont{and}
		\bibinfo{author}{\bibfnamefont{F.}~\bibnamefont{Schmid}}, in
		\emph{\bibinfo{booktitle}{From Single Molecules to Nanoscopically Structured
				Materials}}, edited by \bibinfo{editor}{\bibfnamefont{B.}~\bibnamefont{T.}},
		\bibinfo{editor}{\bibfnamefont{M.}~\bibnamefont{K.}}, \bibnamefont{and}
		\bibinfo{editor}{\bibfnamefont{S.}~\bibnamefont{M}}
		(\bibinfo{publisher}{Springer}, \bibinfo{year}{2013}), pp.
		\bibinfo{pages}{237--283}.
		
		\bibitem[{\citenamefont{Coster}(2003)}]{Cos03}
		\bibinfo{author}{\bibfnamefont{H.~G.~L.} \bibnamefont{Coster}},
		\bibinfo{journal}{Journal of Biological Physics}
		\textbf{\bibinfo{volume}{29}}, \bibinfo{pages}{363} (\bibinfo{year}{2003}).
		
		\bibitem[{\citenamefont{Perlmutter and Hagan}(2015{\natexlab{a}})}]{Jas15}
		\bibinfo{author}{\bibfnamefont{J.~D.} \bibnamefont{Perlmutter}}
		\bibnamefont{and} \bibinfo{author}{\bibfnamefont{M.~F.} \bibnamefont{Hagan}},
		\bibinfo{journal}{Annual Review of Physical Chemistry}
		\textbf{\bibinfo{volume}{66}}, \bibinfo{pages}{217}
		(\bibinfo{year}{2015}{\natexlab{a}}).
		
		\bibitem[{\citenamefont{Hu et~al.}(2015)\citenamefont{Hu, Stanzione, Sum,
				Faller, and Deserno}}]{Hu15}
		\bibinfo{author}{\bibfnamefont{M.}~\bibnamefont{Hu}},
		\bibinfo{author}{\bibfnamefont{F.}~\bibnamefont{Stanzione}},
		\bibinfo{author}{\bibfnamefont{A.~K.} \bibnamefont{Sum}},
		\bibinfo{author}{\bibfnamefont{R.}~\bibnamefont{Faller}}, \bibnamefont{and}
		\bibinfo{author}{\bibfnamefont{M.}~\bibnamefont{Deserno}},
		\bibinfo{journal}{ACS Nano} \textbf{\bibinfo{volume}{9}},
		\bibinfo{pages}{9942} (\bibinfo{year}{2015}).
		
		\bibitem[{\citenamefont{Bernardino de~la Serna
				et~al.}(2016)\citenamefont{Bernardino de~la Serna, Schütz, Eggeling, and
				Cebecauer}}]{Ser16}
		\bibinfo{author}{\bibfnamefont{J.}~\bibnamefont{Bernardino de~la Serna}},
		\bibinfo{author}{\bibfnamefont{G.~J.} \bibnamefont{Schütz}},
		\bibinfo{author}{\bibfnamefont{C.}~\bibnamefont{Eggeling}}, \bibnamefont{and}
		\bibinfo{author}{\bibfnamefont{M.}~\bibnamefont{Cebecauer}},
		\bibinfo{journal}{Frontiers in Cell and Developmental Biology}
		\textbf{\bibinfo{volume}{4}}, \bibinfo{pages}{106} (\bibinfo{year}{2016}).
		
		\bibitem[{\citenamefont{Deserno and Gelbart}(2002)}]{Mar02}
		\bibinfo{author}{\bibfnamefont{M.}~\bibnamefont{Deserno}} \bibnamefont{and}
		\bibinfo{author}{\bibfnamefont{W.~M.} \bibnamefont{Gelbart}},
		\bibinfo{journal}{The Journal of Physical Chemistry B}
		\textbf{\bibinfo{volume}{106}}, \bibinfo{pages}{5543} (\bibinfo{year}{2002}).
		
		\bibitem[{\citenamefont{Murtola et~al.}(2009)\citenamefont{Murtola, Bunker,
				Vattulainen, Deserno, and Karttunen}}]{Mur09}
		\bibinfo{author}{\bibfnamefont{T.}~\bibnamefont{Murtola}},
		\bibinfo{author}{\bibfnamefont{A.}~\bibnamefont{Bunker}},
		\bibinfo{author}{\bibfnamefont{I.}~\bibnamefont{Vattulainen}},
		\bibinfo{author}{\bibfnamefont{M.}~\bibnamefont{Deserno}}, \bibnamefont{and}
		\bibinfo{author}{\bibfnamefont{M.}~\bibnamefont{Karttunen}},
		\bibinfo{journal}{Phys. Chem. Chem. Phys.} \textbf{\bibinfo{volume}{11}},
		\bibinfo{pages}{1869} (\bibinfo{year}{2009}).
		
		\bibitem[{\citenamefont{van Rijn et~al.}(2013)\citenamefont{van Rijn, Tutus,
				Kathrein, Zhu, Wessling, Schwaneberg, and Böker}}]{Rij13}
		\bibinfo{author}{\bibfnamefont{P.}~\bibnamefont{van Rijn}},
		\bibinfo{author}{\bibfnamefont{M.}~\bibnamefont{Tutus}},
		\bibinfo{author}{\bibfnamefont{C.}~\bibnamefont{Kathrein}},
		\bibinfo{author}{\bibfnamefont{L.}~\bibnamefont{Zhu}},
		\bibinfo{author}{\bibfnamefont{M.}~\bibnamefont{Wessling}},
		\bibinfo{author}{\bibfnamefont{U.}~\bibnamefont{Schwaneberg}},
		\bibnamefont{and} \bibinfo{author}{\bibfnamefont{A.}~\bibnamefont{Böker}},
		\bibinfo{journal}{Chem. Soc. Rev.} \textbf{\bibinfo{volume}{42}},
		\bibinfo{pages}{6578} (\bibinfo{year}{2013}).
		
		\bibitem[{\citenamefont{Matthews and Likos}(2013{\natexlab{a}})}]{Matt13}
		\bibinfo{author}{\bibfnamefont{R.}~\bibnamefont{Matthews}} \bibnamefont{and}
		\bibinfo{author}{\bibfnamefont{C.~N.} \bibnamefont{Likos}},
		\bibinfo{journal}{Soft Matter} \textbf{\bibinfo{volume}{9}},
		\bibinfo{pages}{5794} (\bibinfo{year}{2013}{\natexlab{a}}).
		
		\bibitem[{\citenamefont{Wu et~al.}(2014)\citenamefont{Wu, Sheng, and
				Tsao}}]{Wu14}
		\bibinfo{author}{\bibfnamefont{H.-L.} \bibnamefont{Wu}},
		\bibinfo{author}{\bibfnamefont{Y.-J.} \bibnamefont{Sheng}}, \bibnamefont{and}
		\bibinfo{author}{\bibfnamefont{H.-K.} \bibnamefont{Tsao}},
		\bibinfo{journal}{The Journal of Chemical Physics}
		\textbf{\bibinfo{volume}{141}}, \bibinfo{pages}{124906}
		(\bibinfo{year}{2014}).
		
		\bibitem[{\citenamefont{Boles et~al.}(2016)\citenamefont{Boles, Engel, and
				Talapin}}]{Bol16}
		\bibinfo{author}{\bibfnamefont{M.~A.} \bibnamefont{Boles}},
		\bibinfo{author}{\bibfnamefont{M.}~\bibnamefont{Engel}}, \bibnamefont{and}
		\bibinfo{author}{\bibfnamefont{D.~V.} \bibnamefont{Talapin}},
		\bibinfo{journal}{Chemical Reviews} \textbf{\bibinfo{volume}{116}},
		\bibinfo{pages}{11220} (\bibinfo{year}{2016}).
		
		\bibitem[{\citenamefont{Deserno}(2009)}]{Des09}
		\bibinfo{author}{\bibfnamefont{M.}~\bibnamefont{Deserno}},
		\bibinfo{journal}{Macromolecular Rapid Communications}
		\textbf{\bibinfo{volume}{30}}, \bibinfo{pages}{752} (\bibinfo{year}{2009}).
		
		\bibitem[{\citenamefont{Noguchi}(2013)}]{Nog13}
		\bibinfo{author}{\bibfnamefont{H.}~\bibnamefont{Noguchi}},
		\bibinfo{journal}{AIP Conference Proceedings}
		\textbf{\bibinfo{volume}{1518}}, \bibinfo{pages}{566} (\bibinfo{year}{2013}).
		
		\bibitem[{\citenamefont{Perlmutter and Hagan}(2015{\natexlab{b}})}]{Per15}
		\bibinfo{author}{\bibfnamefont{J.~D.} \bibnamefont{Perlmutter}}
		\bibnamefont{and} \bibinfo{author}{\bibfnamefont{M.~F.} \bibnamefont{Hagan}},
		\bibinfo{journal}{Annual Review of Physical Chemistry}
		\textbf{\bibinfo{volume}{66}}, \bibinfo{pages}{217}
		(\bibinfo{year}{2015}{\natexlab{b}}).
		
		\bibitem[{\citenamefont{Khosravanizadeh
				et~al.}(2019)\citenamefont{Khosravanizadeh, Sens, and
				Mohammad-Rafiee}}]{Kho19}
		\bibinfo{author}{\bibfnamefont{A.}~\bibnamefont{Khosravanizadeh}},
		\bibinfo{author}{\bibfnamefont{P.}~\bibnamefont{Sens}}, \bibnamefont{and}
		\bibinfo{author}{\bibfnamefont{F.}~\bibnamefont{Mohammad-Rafiee}},
		\bibinfo{journal}{Soft Matter} \textbf{\bibinfo{volume}{15}},
		\bibinfo{pages}{7490} (\bibinfo{year}{2019}).
		
		\bibitem[{\citenamefont{Holowka et~al.}(2005)\citenamefont{Holowka, Pochan, and
				Deming}}]{Hol05}
		\bibinfo{author}{\bibfnamefont{E.~P.} \bibnamefont{Holowka}},
		\bibinfo{author}{\bibfnamefont{D.~J.} \bibnamefont{Pochan}},
		\bibnamefont{and} \bibinfo{author}{\bibfnamefont{T.~J.}
			\bibnamefont{Deming}}, \bibinfo{journal}{Journal of the American Chemical
			Society} \textbf{\bibinfo{volume}{127}}, \bibinfo{pages}{12423}
		(\bibinfo{year}{2005}).
		
		\bibitem[{\citenamefont{Fatouros et~al.}(2014)\citenamefont{Fatouros, Lamprou,
				Urquhart, Yannopoulos, Vizirianakis, Zhang, and Koutsopoulos}}]{Fat14}
		\bibinfo{author}{\bibfnamefont{D.~G.} \bibnamefont{Fatouros}},
		\bibinfo{author}{\bibfnamefont{D.~A.} \bibnamefont{Lamprou}},
		\bibinfo{author}{\bibfnamefont{A.~J.} \bibnamefont{Urquhart}},
		\bibinfo{author}{\bibfnamefont{S.~N.} \bibnamefont{Yannopoulos}},
		\bibinfo{author}{\bibfnamefont{I.~S.} \bibnamefont{Vizirianakis}},
		\bibinfo{author}{\bibfnamefont{S.}~\bibnamefont{Zhang}}, \bibnamefont{and}
		\bibinfo{author}{\bibfnamefont{S.}~\bibnamefont{Koutsopoulos}},
		\bibinfo{journal}{ACS Applied Materials \& Interfaces}
		\textbf{\bibinfo{volume}{6}}, \bibinfo{pages}{8184} (\bibinfo{year}{2014}).
		
		\bibitem[{\citenamefont{Thanuja et~al.}(2018)\citenamefont{Thanuja, Anupama,
				and Sudhir}}]{Tha18}
		\bibinfo{author}{\bibfnamefont{M.}~\bibnamefont{Thanuja}},
		\bibinfo{author}{\bibfnamefont{C.}~\bibnamefont{Anupama}}, \bibnamefont{and}
		\bibinfo{author}{\bibfnamefont{H.~R.} \bibnamefont{Sudhir}},
		\bibinfo{journal}{Advanced Drug Delivery Reviews}
		\textbf{\bibinfo{volume}{132}}, \bibinfo{pages}{57 } (\bibinfo{year}{2018}).
		
		\bibitem[{\citenamefont{Pandit et~al.}(2018)\citenamefont{Pandit, Roy, Agarwal,
				and Chatterjee}}]{Pan18}
		\bibinfo{author}{\bibfnamefont{G.}~\bibnamefont{Pandit}},
		\bibinfo{author}{\bibfnamefont{K.}~\bibnamefont{Roy}},
		\bibinfo{author}{\bibfnamefont{U.}~\bibnamefont{Agarwal}}, \bibnamefont{and}
		\bibinfo{author}{\bibfnamefont{S.}~\bibnamefont{Chatterjee}},
		\bibinfo{journal}{ACS Omega} \textbf{\bibinfo{volume}{3}},
		\bibinfo{pages}{3143} (\bibinfo{year}{2018}).
		
		\bibitem[{\citenamefont{Daddi-Moussa-Ider
				et~al.}(2019)\citenamefont{Daddi-Moussa-Ider, Goh, Liebchen, Hoell,
				Mathijssen, Guzm\'an-Lastra, Scholz, Menzel, and L\"owen}}]{Dad19}
		\bibinfo{author}{\bibfnamefont{A.}~\bibnamefont{Daddi-Moussa-Ider}},
		\bibinfo{author}{\bibfnamefont{S.}~\bibnamefont{Goh}},
		\bibinfo{author}{\bibfnamefont{B.}~\bibnamefont{Liebchen}},
		\bibinfo{author}{\bibfnamefont{C.}~\bibnamefont{Hoell}},
		\bibinfo{author}{\bibfnamefont{A.~J. T.~M.} \bibnamefont{Mathijssen}},
		\bibinfo{author}{\bibfnamefont{F.}~\bibnamefont{Guzm\'an-Lastra}},
		\bibinfo{author}{\bibfnamefont{C.}~\bibnamefont{Scholz}},
		\bibinfo{author}{\bibfnamefont{A.~M.} \bibnamefont{Menzel}},
		\bibnamefont{and} \bibinfo{author}{\bibfnamefont{H.}~\bibnamefont{L\"owen}},
		\bibinfo{journal}{The Journal of Chemical Physics}
		\textbf{\bibinfo{volume}{150}}, \bibinfo{pages}{064906}
		(\bibinfo{year}{2019}).
		
		\bibitem[{\citenamefont{Gelbart and Knobler}(2008)}]{Gel08}
		\bibinfo{author}{\bibfnamefont{M.~W.} \bibnamefont{Gelbart}} \bibnamefont{and}
		\bibinfo{author}{\bibfnamefont{M.~C.} \bibnamefont{Knobler}},
		\bibinfo{journal}{Physics Today} \textbf{\bibinfo{volume}{61}},
		\bibinfo{pages}{42} (\bibinfo{year}{2008}).
		
		\bibitem[{\citenamefont{Schmid}(2017)}]{Schm17}
		\bibinfo{author}{\bibfnamefont{F.}~\bibnamefont{Schmid}},
		\bibinfo{journal}{Biochimica et biophysica acta. Biomembranes}
		\textbf{\bibinfo{volume}{1859 4}}, \bibinfo{pages}{509}
		(\bibinfo{year}{2017}).
		
		\bibitem[{\citenamefont{Chand et~al.}(2019)\citenamefont{Chand, Beales,
				Claeyssens, and Ciani}}]{Cha19}
		\bibinfo{author}{\bibfnamefont{S.}~\bibnamefont{Chand}},
		\bibinfo{author}{\bibfnamefont{P.}~\bibnamefont{Beales}},
		\bibinfo{author}{\bibfnamefont{F.}~\bibnamefont{Claeyssens}},
		\bibnamefont{and} \bibinfo{author}{\bibfnamefont{B.}~\bibnamefont{Ciani}},
		\bibinfo{journal}{Experimental Biology and Medicine}
		\textbf{\bibinfo{volume}{244}}, \bibinfo{pages}{294} (\bibinfo{year}{2019}).
		
		\bibitem[{\citenamefont{Gerle}(2019)}]{Ger19}
		\bibinfo{author}{\bibfnamefont{C.}~\bibnamefont{Gerle}}, \bibinfo{journal}{The
			Journal of Membrane Biology} \textbf{\bibinfo{volume}{252}},
		\bibinfo{pages}{115} (\bibinfo{year}{2019}).
		
		\bibitem[{\citenamefont{Levin et~al.}(2004)\citenamefont{Levin, Idiart, and
				Arenzon}}]{Lev04}
		\bibinfo{author}{\bibfnamefont{Y.}~\bibnamefont{Levin}},
		\bibinfo{author}{\bibfnamefont{M.~A.} \bibnamefont{Idiart}},
		\bibnamefont{and} \bibinfo{author}{\bibfnamefont{J.~J.}
			\bibnamefont{Arenzon}}, \bibinfo{journal}{Physica A: Statistical Mechanics
			and its Applications} \textbf{\bibinfo{volume}{344}}, \bibinfo{pages}{543 }
		(\bibinfo{year}{2004}).
		
		\bibitem[{\citenamefont{Idiart and Levin}(2004)}]{Idi04}
		\bibinfo{author}{\bibfnamefont{M.~A.} \bibnamefont{Idiart}} \bibnamefont{and}
		\bibinfo{author}{\bibfnamefont{Y.}~\bibnamefont{Levin}},
		\bibinfo{journal}{Phys. Rev. E} \textbf{\bibinfo{volume}{69}},
		\bibinfo{pages}{061922} (\bibinfo{year}{2004}).
		
		\bibitem[{\citenamefont{Levin and Idiart}(2004)}]{Levin04}
		\bibinfo{author}{\bibfnamefont{Y.}~\bibnamefont{Levin}} \bibnamefont{and}
		\bibinfo{author}{\bibfnamefont{M.~A.} \bibnamefont{Idiart}},
		\bibinfo{journal}{Physica A: Statistical Mechanics and its Applications}
		\textbf{\bibinfo{volume}{331}}, \bibinfo{pages}{571 } (\bibinfo{year}{2004}).
		
		\bibitem[{\citenamefont{Cooke and Deserno}(2006)}]{Coo06}
		\bibinfo{author}{\bibfnamefont{I.~R.} \bibnamefont{Cooke}} \bibnamefont{and}
		\bibinfo{author}{\bibfnamefont{M.}~\bibnamefont{Deserno}},
		\bibinfo{journal}{Biophysical journal} \textbf{\bibinfo{volume}{91}},
		\bibinfo{pages}{487} (\bibinfo{year}{2006}).
		
		\bibitem[{\citenamefont{Deserno}(2015)}]{Des15}
		\bibinfo{author}{\bibfnamefont{M.}~\bibnamefont{Deserno}},
		\bibinfo{journal}{Chemistry and Physics of Lipids}
		\textbf{\bibinfo{volume}{185}}, \bibinfo{pages}{11 } (\bibinfo{year}{2015}).
		
		\bibitem[{\citenamefont{Mateu}(2013)}]{Mate13}
		\bibinfo{author}{\bibfnamefont{M.~G.} \bibnamefont{Mateu}},
		\bibinfo{journal}{Archives of Biochemistry and Biophysics}
		\textbf{\bibinfo{volume}{531}}, \bibinfo{pages}{65 } (\bibinfo{year}{2013}).
		
		\bibitem[{\citenamefont{Lozada-Cassou and Yu}(1996)}]{Loz96}
		\bibinfo{author}{\bibfnamefont{M.}~\bibnamefont{Lozada-Cassou}}
		\bibnamefont{and} \bibinfo{author}{\bibfnamefont{J.}~\bibnamefont{Yu}},
		\bibinfo{journal}{Phys. Rev. Lett.} \textbf{\bibinfo{volume}{77}},
		\bibinfo{pages}{4019} (\bibinfo{year}{1996}).
		
		\bibitem[{\citenamefont{Levin}(2002)}]{Lev02}
		\bibinfo{author}{\bibfnamefont{Y.}~\bibnamefont{Levin}}, \bibinfo{journal}{Rep.
			Prog. Phys} \textbf{\bibinfo{volume}{65}}, \bibinfo{pages}{1577}
		(\bibinfo{year}{2002}).
		
		\bibitem[{\citenamefont{Lozada-Cassou}(1992)}]{Lozada}
		\bibinfo{author}{\bibfnamefont{M.}~\bibnamefont{Lozada-Cassou}}, in
		\emph{\bibinfo{booktitle}{Fundamentals of Inhomogeneous Fluids}}, edited by
		\bibinfo{editor}{\bibfnamefont{D.}~\bibnamefont{Henderson}}
		(\bibinfo{publisher}{Marcel Dekker, New York}, \bibinfo{year}{1992}),
		chap.~\bibinfo{chapter}{8}.
		
		\bibitem[{\citenamefont{Zandi and Reguera}(2005)}]{Zan05}
		\bibinfo{author}{\bibfnamefont{R.}~\bibnamefont{Zandi}} \bibnamefont{and}
		\bibinfo{author}{\bibfnamefont{D.}~\bibnamefont{Reguera}},
		\bibinfo{journal}{Phys. Rev. E} \textbf{\bibinfo{volume}{72}},
		\bibinfo{pages}{021917} (\bibinfo{year}{2005}).
		
		\bibitem[{\citenamefont{Nguyen et~al.}(2005)\citenamefont{Nguyen, Bruinsma, and
				Gelbart}}]{Ngu05}
		\bibinfo{author}{\bibfnamefont{T.~T.} \bibnamefont{Nguyen}},
		\bibinfo{author}{\bibfnamefont{R.~F.} \bibnamefont{Bruinsma}},
		\bibnamefont{and} \bibinfo{author}{\bibfnamefont{W.~M.}
			\bibnamefont{Gelbart}}, \bibinfo{journal}{Phys. Rev. E}
		\textbf{\bibinfo{volume}{72}}, \bibinfo{pages}{051923}
		(\bibinfo{year}{2005}).
		
		\bibitem[{\citenamefont{Nguyen et~al.}(2006)\citenamefont{Nguyen, Bruinsma, and
				Gelbart}}]{Ngu06}
		\bibinfo{author}{\bibfnamefont{T.~T.} \bibnamefont{Nguyen}},
		\bibinfo{author}{\bibfnamefont{R.~F.} \bibnamefont{Bruinsma}},
		\bibnamefont{and} \bibinfo{author}{\bibfnamefont{W.~M.}
			\bibnamefont{Gelbart}}, \bibinfo{journal}{Phys. Rev. Lett.}
		\textbf{\bibinfo{volume}{96}}, \bibinfo{pages}{078102}
		(\bibinfo{year}{2006}).
		
		\bibitem[{\citenamefont{Michel et~al.}(2006)\citenamefont{Michel, Ivanovska,
				Gibbons, Klug, Knobler, Wuite, and Schmidt}}]{Mic06}
		\bibinfo{author}{\bibfnamefont{J.~P.} \bibnamefont{Michel}},
		\bibinfo{author}{\bibfnamefont{I.~L.} \bibnamefont{Ivanovska}},
		\bibinfo{author}{\bibfnamefont{M.~M.} \bibnamefont{Gibbons}},
		\bibinfo{author}{\bibfnamefont{W.~S.} \bibnamefont{Klug}},
		\bibinfo{author}{\bibfnamefont{C.~M.} \bibnamefont{Knobler}},
		\bibinfo{author}{\bibfnamefont{G.~J.~L.} \bibnamefont{Wuite}},
		\bibnamefont{and} \bibinfo{author}{\bibfnamefont{C.~F.}
			\bibnamefont{Schmidt}}, \bibinfo{journal}{Proceedings of the National Academy
			of Sciences} \textbf{\bibinfo{volume}{103}}, \bibinfo{pages}{6184}
		(\bibinfo{year}{2006}).
		
		\bibitem[{\citenamefont{Hu et~al.}(2013)\citenamefont{Hu, de~Jong, Marrink, and
				Deserno}}]{Hu13}
		\bibinfo{author}{\bibfnamefont{M.}~\bibnamefont{Hu}},
		\bibinfo{author}{\bibfnamefont{D.~H.} \bibnamefont{de~Jong}},
		\bibinfo{author}{\bibfnamefont{S.~J.} \bibnamefont{Marrink}},
		\bibnamefont{and} \bibinfo{author}{\bibfnamefont{M.}~\bibnamefont{Deserno}},
		\bibinfo{journal}{Faraday Discuss.} \textbf{\bibinfo{volume}{161}},
		\bibinfo{pages}{365} (\bibinfo{year}{2013}).
		
		\bibitem[{\citenamefont{Krishnamani et~al.}(2016)\citenamefont{Krishnamani,
				Globisch, Peter, and Deserno}}]{Kri16}
		\bibinfo{author}{\bibfnamefont{V.}~\bibnamefont{Krishnamani}},
		\bibinfo{author}{\bibfnamefont{C.}~\bibnamefont{Globisch}},
		\bibinfo{author}{\bibfnamefont{C.}~\bibnamefont{Peter}}, \bibnamefont{and}
		\bibinfo{author}{\bibfnamefont{M.}~\bibnamefont{Deserno}},
		\bibinfo{journal}{The European Physical Journal Special Topics}
		\textbf{\bibinfo{volume}{225}}, \bibinfo{pages}{1317–1321}
		(\bibinfo{year}{2016}).
		
		\bibitem[{\citenamefont{Norouzi et~al.}(2006)\citenamefont{Norouzi, M\"uller,
				and Deserno}}]{Nor06}
		\bibinfo{author}{\bibfnamefont{D.}~\bibnamefont{Norouzi}},
		\bibinfo{author}{\bibfnamefont{M.~M.} \bibnamefont{M\"uller}},
		\bibnamefont{and} \bibinfo{author}{\bibfnamefont{M.}~\bibnamefont{Deserno}},
		\bibinfo{journal}{Phys. Rev. E} \textbf{\bibinfo{volume}{74}},
		\bibinfo{pages}{061914} (\bibinfo{year}{2006}).
		
		\bibitem[{\citenamefont{Mouritsen}(1987)}]{Mou87}
		\bibinfo{author}{\bibfnamefont{O.~G.} \bibnamefont{Mouritsen}}, in
		\emph{\bibinfo{booktitle}{Physics in Living Matter}}, edited by
		\bibinfo{editor}{\bibfnamefont{B.}~\bibnamefont{D.}},
		\bibinfo{editor}{\bibfnamefont{D.}~\bibnamefont{M.}},
		\bibinfo{editor}{\bibfnamefont{M.}~\bibnamefont{A.}}, \bibnamefont{and}
		\bibinfo{editor}{\bibfnamefont{M.}~\bibnamefont{P.}}
		(\bibinfo{publisher}{Springer, Berlin, Heidelberg}, \bibinfo{year}{1987}),
		vol. \bibinfo{volume}{284}.
		
		\bibitem[{\citenamefont{Javidpour et~al.}(2019)\citenamefont{Javidpour,
				Lo\v{s}dorfer~Bo\v{z}i\v{c}, Podgornik, and Naji}}]{Jav19}
		\bibinfo{author}{\bibfnamefont{L.}~\bibnamefont{Javidpour}},
		\bibinfo{author}{\bibfnamefont{A.}~\bibnamefont{Lo\v{s}dorfer~Bo\v{z}i\v{c}}},
		\bibinfo{author}{\bibfnamefont{R.}~\bibnamefont{Podgornik}},
		\bibnamefont{and} \bibinfo{author}{\bibfnamefont{A.}~\bibnamefont{Naji}},
		\bibinfo{journal}{Scientific Reports} \textbf{\bibinfo{volume}{9}},
		\bibinfo{pages}{3884} (\bibinfo{year}{2019}).
		
		\bibitem[{\citenamefont{\v{S}iber et~al.}(2012)\citenamefont{\v{S}iber,
				Bo\v{z}i\v{c}, and Podgornik}}]{Sil12}
		\bibinfo{author}{\bibfnamefont{A.}~\bibnamefont{\v{S}iber}},
		\bibinfo{author}{\bibfnamefont{A.~L.} \bibnamefont{Bo\v{z}i\v{c}}},
		\bibnamefont{and}
		\bibinfo{author}{\bibfnamefont{R.}~\bibnamefont{Podgornik}},
		\bibinfo{journal}{Phys. Chem. Chem. Phys.} \textbf{\bibinfo{volume}{14}},
		\bibinfo{pages}{3746} (\bibinfo{year}{2012}).
		
		\bibitem[{\citenamefont{Hernando-P\'erez
				et~al.}(2015)\citenamefont{Hernando-P\'erez, Cartagena-Rivera,
				Lošdorfer~Božič, Carrillo, San~Mart\'in, Mateu, Raman, Podgornik, and
				de~Pablo}}]{Her15}
		\bibinfo{author}{\bibfnamefont{M.}~\bibnamefont{Hernando-P\'erez}},
		\bibinfo{author}{\bibfnamefont{A.~X.} \bibnamefont{Cartagena-Rivera}},
		\bibinfo{author}{\bibfnamefont{A.}~\bibnamefont{Lošdorfer~Božič}},
		\bibinfo{author}{\bibfnamefont{P.~J.~P.} \bibnamefont{Carrillo}},
		\bibinfo{author}{\bibfnamefont{C.}~\bibnamefont{San~Mart\'in}},
		\bibinfo{author}{\bibfnamefont{M.~G.} \bibnamefont{Mateu}},
		\bibinfo{author}{\bibfnamefont{A.}~\bibnamefont{Raman}},
		\bibinfo{author}{\bibfnamefont{R.}~\bibnamefont{Podgornik}},
		\bibnamefont{and} \bibinfo{author}{\bibfnamefont{P.~J.}
			\bibnamefont{de~Pablo}}, \bibinfo{journal}{Nanoscale}
		\textbf{\bibinfo{volume}{7}}, \bibinfo{pages}{17289} (\bibinfo{year}{2015}).
		
		\bibitem[{\citenamefont{Shojaei et~al.}(2016)\citenamefont{Shojaei, Bo\ifmmode
				\check{z}\else \v{z}\fi{}i\ifmmode~\check{c}\else \v{c}\fi{}, Muthukumar, and
				Podgornik}}]{Sho16}
		\bibinfo{author}{\bibfnamefont{H.~R.} \bibnamefont{Shojaei}},
		\bibinfo{author}{\bibfnamefont{A.~c. v. L. c.~v.} \bibnamefont{Bo\ifmmode
				\check{z}\else \v{z}\fi{}i\ifmmode~\check{c}\else \v{c}\fi{}}},
		\bibinfo{author}{\bibfnamefont{M.}~\bibnamefont{Muthukumar}},
		\bibnamefont{and}
		\bibinfo{author}{\bibfnamefont{R.}~\bibnamefont{Podgornik}},
		\bibinfo{journal}{Phys. Rev. E} \textbf{\bibinfo{volume}{93}},
		\bibinfo{pages}{052415} (\bibinfo{year}{2016}).
		
		\bibitem[{\citenamefont{Ko\v{s}mrlj and Nelson}(2017)}]{Kov17}
		\bibinfo{author}{\bibfnamefont{A.}~\bibnamefont{Ko\v{s}mrlj}} \bibnamefont{and}
		\bibinfo{author}{\bibfnamefont{D.~R.} \bibnamefont{Nelson}},
		\bibinfo{journal}{Phys. Rev. X} \textbf{\bibinfo{volume}{7}},
		\bibinfo{pages}{011002} (\bibinfo{year}{2017}).
		
		\bibitem[{\citenamefont{Sun et~al.}(2018)\citenamefont{Sun, Li, Wang, Yin, Hu,
				Li, Liu, Gao, Ren, Zheng et~al.}}]{Sun18}
		\bibinfo{author}{\bibfnamefont{X.}~\bibnamefont{Sun}},
		\bibinfo{author}{\bibfnamefont{D.}~\bibnamefont{Li}},
		\bibinfo{author}{\bibfnamefont{Z.}~\bibnamefont{Wang}},
		\bibinfo{author}{\bibfnamefont{P.}~\bibnamefont{Yin}},
		\bibinfo{author}{\bibfnamefont{R.}~\bibnamefont{Hu}},
		\bibinfo{author}{\bibfnamefont{H.}~\bibnamefont{Li}},
		\bibinfo{author}{\bibfnamefont{Q.}~\bibnamefont{Liu}},
		\bibinfo{author}{\bibfnamefont{Y.}~\bibnamefont{Gao}},
		\bibinfo{author}{\bibfnamefont{B.}~\bibnamefont{Ren}},
		\bibinfo{author}{\bibfnamefont{J.}~\bibnamefont{Zheng}},
		\bibnamefont{et~al.}, \bibinfo{journal}{ACS Omega}
		\textbf{\bibinfo{volume}{3}}, \bibinfo{pages}{4384} (\bibinfo{year}{2018}).
		
		\bibitem[{\citenamefont{Bo\v{z}i\v{c} and \v{S}iber}(2018)}]{Anv18}
		\bibinfo{author}{\bibfnamefont{A.~L.} \bibnamefont{Bo\v{z}i\v{c}}}
		\bibnamefont{and}
		\bibinfo{author}{\bibfnamefont{A.}~\bibnamefont{\v{S}iber}},
		\bibinfo{journal}{Biophysical Journal} \textbf{\bibinfo{volume}{115}},
		\bibinfo{pages}{822 } (\bibinfo{year}{2018}).
		
		\bibitem[{\citenamefont{Reguera et~al.}(2019)\citenamefont{Reguera,
				Hernández-Rojas, and Gomez~Llorente}}]{Reg19}
		\bibinfo{author}{\bibfnamefont{D.}~\bibnamefont{Reguera}},
		\bibinfo{author}{\bibfnamefont{J.}~\bibnamefont{Hernández-Rojas}},
		\bibnamefont{and} \bibinfo{author}{\bibfnamefont{J.~M.}
			\bibnamefont{Gomez~Llorente}}, \bibinfo{journal}{Soft Matter}
		\textbf{\bibinfo{volume}{15}}, \bibinfo{pages}{7166} (\bibinfo{year}{2019}).
		
		\bibitem[{\citenamefont{Xian et~al.}(2019)\citenamefont{Xian, Karki, Silva, Li,
				and Xiao}}]{Xia19}
		\bibinfo{author}{\bibfnamefont{Y.}~\bibnamefont{Xian}},
		\bibinfo{author}{\bibfnamefont{C.~B.} \bibnamefont{Karki}},
		\bibinfo{author}{\bibfnamefont{S.~M.} \bibnamefont{Silva}},
		\bibinfo{author}{\bibfnamefont{L.}~\bibnamefont{Li}}, \bibnamefont{and}
		\bibinfo{author}{\bibfnamefont{C.}~\bibnamefont{Xiao}},
		\bibinfo{journal}{International journal of molecular sciences}
		\textbf{\bibinfo{volume}{20}}, \bibinfo{pages}{1876} (\bibinfo{year}{2019}).
		
		\bibitem[{\citenamefont{Almendral}(2013)}]{Almendral}
		\bibinfo{author}{\bibfnamefont{J.~M.} \bibnamefont{Almendral}}, in
		\emph{\bibinfo{booktitle}{Structure and Physics of Viruses}}, edited by
		\bibinfo{editor}{\bibfnamefont{M.}~\bibnamefont{Mateu}}
		(\bibinfo{publisher}{Springer, Dordrecht}, \bibinfo{year}{2013}), pp.
		\bibinfo{pages}{307--328}.
		
		\bibitem[{\citenamefont{Roos et~al.}(2007)\citenamefont{Roos, Ivanovska,
				Evilevitch, and Wuite}}]{Roo07}
		\bibinfo{author}{\bibfnamefont{W.~H.} \bibnamefont{Roos}},
		\bibinfo{author}{\bibfnamefont{I.~L.} \bibnamefont{Ivanovska}},
		\bibinfo{author}{\bibfnamefont{A.}~\bibnamefont{Evilevitch}},
		\bibnamefont{and} \bibinfo{author}{\bibfnamefont{G.~J.} \bibnamefont{Wuite}},
		\bibinfo{journal}{Cellular and molecular life sciences : CMLS}
		\textbf{\bibinfo{volume}{64}}, \bibinfo{pages}{1484–1497}
		(\bibinfo{year}{2007}).
		
		\bibitem[{\citenamefont{Johnston et~al.}(2010)\citenamefont{Johnston, Louis,
				and Doye}}]{Joh10}
		\bibinfo{author}{\bibfnamefont{I.~G.} \bibnamefont{Johnston}},
		\bibinfo{author}{\bibfnamefont{A.~A.} \bibnamefont{Louis}}, \bibnamefont{and}
		\bibinfo{author}{\bibfnamefont{J.~P.~K.} \bibnamefont{Doye}},
		\bibinfo{journal}{Journal of Physics: Condensed Matter}
		\textbf{\bibinfo{volume}{22}}, \bibinfo{pages}{104101}
		(\bibinfo{year}{2010}).
		
		\bibitem[{\citenamefont{Matthews and Likos}(2013{\natexlab{b}})}]{Mat13}
		\bibinfo{author}{\bibfnamefont{R.}~\bibnamefont{Matthews}} \bibnamefont{and}
		\bibinfo{author}{\bibfnamefont{C.~N.} \bibnamefont{Likos}},
		\bibinfo{journal}{The Journal of Physical Chemistry B}
		\textbf{\bibinfo{volume}{117}}, \bibinfo{pages}{8283}
		(\bibinfo{year}{2013}{\natexlab{b}}).
		
		\bibitem[{\citenamefont{Comas-Garcia et~al.}(2014)\citenamefont{Comas-Garcia,
				Garmann, Singaram, Ben-Shaul, Knobler, and Gelbart}}]{Com14}
		\bibinfo{author}{\bibfnamefont{M.}~\bibnamefont{Comas-Garcia}},
		\bibinfo{author}{\bibfnamefont{R.~F.} \bibnamefont{Garmann}},
		\bibinfo{author}{\bibfnamefont{S.~W.} \bibnamefont{Singaram}},
		\bibinfo{author}{\bibfnamefont{A.}~\bibnamefont{Ben-Shaul}},
		\bibinfo{author}{\bibfnamefont{C.~M.} \bibnamefont{Knobler}},
		\bibnamefont{and} \bibinfo{author}{\bibfnamefont{W.~M.}
			\bibnamefont{Gelbart}}, \bibinfo{journal}{The Journal of Physical Chemistry
			B} \textbf{\bibinfo{volume}{118}}, \bibinfo{pages}{7510}
		(\bibinfo{year}{2014}).
		
		\bibitem[{\citenamefont{Bruinsma et~al.}(2016)\citenamefont{Bruinsma,
				Comas-Garcia, Garmann, and Grosberg}}]{Bru16}
		\bibinfo{author}{\bibfnamefont{R.~F.} \bibnamefont{Bruinsma}},
		\bibinfo{author}{\bibfnamefont{M.}~\bibnamefont{Comas-Garcia}},
		\bibinfo{author}{\bibfnamefont{R.~F.} \bibnamefont{Garmann}},
		\bibnamefont{and} \bibinfo{author}{\bibfnamefont{A.~Y.}
			\bibnamefont{Grosberg}}, \bibinfo{journal}{Phys. Rev. E}
		\textbf{\bibinfo{volume}{93}}, \bibinfo{pages}{032405}
		(\bibinfo{year}{2016}).
		
		\bibitem[{\citenamefont{\v{S}iber}(2006)}]{Sil06}
		\bibinfo{author}{\bibfnamefont{A.}~\bibnamefont{\v{S}iber}},
		\bibinfo{journal}{Phys. Rev. E} \textbf{\bibinfo{volume}{73}},
		\bibinfo{pages}{061915} (\bibinfo{year}{2006}).
		
		\bibitem[{\citenamefont{Chen et~al.}(2007)\citenamefont{Chen, Zhang, and
				Glotzer}}]{Che07}
		\bibinfo{author}{\bibfnamefont{T.}~\bibnamefont{Chen}},
		\bibinfo{author}{\bibfnamefont{Z.}~\bibnamefont{Zhang}}, \bibnamefont{and}
		\bibinfo{author}{\bibfnamefont{S.~C.} \bibnamefont{Glotzer}},
		\bibinfo{journal}{Langmuir} \textbf{\bibinfo{volume}{23}},
		\bibinfo{pages}{6598} (\bibinfo{year}{2007}).
		
		\bibitem[{\citenamefont{Aznar et~al.}(2012)\citenamefont{Aznar, Luque, and
				Reguera}}]{Mar12}
		\bibinfo{author}{\bibfnamefont{M.}~\bibnamefont{Aznar}},
		\bibinfo{author}{\bibfnamefont{A.}~\bibnamefont{Luque}}, \bibnamefont{and}
		\bibinfo{author}{\bibfnamefont{D.}~\bibnamefont{Reguera}},
		\bibinfo{journal}{Physical Biology} \textbf{\bibinfo{volume}{9}},
		\bibinfo{pages}{036003} (\bibinfo{year}{2012}).
		
		\bibitem[{\citenamefont{Abrescia et~al.}(2012)\citenamefont{Abrescia, Bamford,
				Grimes, and Stuart}}]{Abr12}
		\bibinfo{author}{\bibfnamefont{N.~G.} \bibnamefont{Abrescia}},
		\bibinfo{author}{\bibfnamefont{D.~H.} \bibnamefont{Bamford}},
		\bibinfo{author}{\bibfnamefont{J.~M.} \bibnamefont{Grimes}},
		\bibnamefont{and} \bibinfo{author}{\bibfnamefont{D.~I.}
			\bibnamefont{Stuart}}, \bibinfo{journal}{Annual Review of Biochemistry}
		\textbf{\bibinfo{volume}{81}}, \bibinfo{pages}{795} (\bibinfo{year}{2012}).
		
		\bibitem[{\citenamefont{Lo\v{s}dorfer Bo\v{z}i\v{c}~A}(2013)}]{Lov13}
		\bibinfo{author}{\bibfnamefont{P.~R.} \bibnamefont{Lo\v{s}dorfer
				Bo\v{z}i\v{c}~A}, \bibfnamefont{\v{S}iber~A}}, \bibinfo{journal}{J. Biol.
			Phys.} \textbf{\bibinfo{volume}{39}}, \bibinfo{pages}{215–228}
		(\bibinfo{year}{2013}).
		
		\bibitem[{\citenamefont{Panahandeh et~al.}(2018)\citenamefont{Panahandeh, Li,
				and Zandi}}]{Pana18}
		\bibinfo{author}{\bibfnamefont{S.}~\bibnamefont{Panahandeh}},
		\bibinfo{author}{\bibfnamefont{S.}~\bibnamefont{Li}}, \bibnamefont{and}
		\bibinfo{author}{\bibfnamefont{R.}~\bibnamefont{Zandi}},
		\bibinfo{journal}{Nanoscale} \textbf{\bibinfo{volume}{10}},
		\bibinfo{pages}{22802} (\bibinfo{year}{2018}).
		
		\bibitem[{\citenamefont{Suci et~al.}(2006)\citenamefont{Suci, Klem, Arce,
				Douglas, and Young}}]{Suc06}
		\bibinfo{author}{\bibfnamefont{P.~A.} \bibnamefont{Suci}},
		\bibinfo{author}{\bibfnamefont{M.~T.} \bibnamefont{Klem}},
		\bibinfo{author}{\bibfnamefont{F.~T.} \bibnamefont{Arce}},
		\bibinfo{author}{\bibfnamefont{T.}~\bibnamefont{Douglas}}, \bibnamefont{and}
		\bibinfo{author}{\bibfnamefont{M.}~\bibnamefont{Young}},
		\bibinfo{journal}{Langmuir} \textbf{\bibinfo{volume}{22}},
		\bibinfo{pages}{8891} (\bibinfo{year}{2006}).
		
		\bibitem[{\citenamefont{Singh et~al.}(2008)\citenamefont{Singh, Al-Jamal,
				Lacerda, and Kostarelos}}]{Sin08}
		\bibinfo{author}{\bibfnamefont{R.}~\bibnamefont{Singh}},
		\bibinfo{author}{\bibfnamefont{K.~T.} \bibnamefont{Al-Jamal}},
		\bibinfo{author}{\bibfnamefont{L.}~\bibnamefont{Lacerda}}, \bibnamefont{and}
		\bibinfo{author}{\bibfnamefont{K.}~\bibnamefont{Kostarelos}},
		\bibinfo{journal}{ACS Nano} \textbf{\bibinfo{volume}{2}},
		\bibinfo{pages}{1040} (\bibinfo{year}{2008}).
		
		\bibitem[{\citenamefont{Costa and Mano}(2014)}]{Cos14}
		\bibinfo{author}{\bibfnamefont{R.~R.} \bibnamefont{Costa}} \bibnamefont{and}
		\bibinfo{author}{\bibfnamefont{J.~a.~F.} \bibnamefont{Mano}},
		\bibinfo{journal}{Chem. Soc. Rev.} \textbf{\bibinfo{volume}{43}},
		\bibinfo{pages}{3453} (\bibinfo{year}{2014}).
		
		\bibitem[{\citenamefont{Miles et~al.}(2015)\citenamefont{Miles, Cassidy,
				Donlon, Yarkoni, and Frankel}}]{Mil15}
		\bibinfo{author}{\bibfnamefont{P.}~\bibnamefont{Miles}},
		\bibinfo{author}{\bibfnamefont{P.}~\bibnamefont{Cassidy}},
		\bibinfo{author}{\bibfnamefont{L.}~\bibnamefont{Donlon}},
		\bibinfo{author}{\bibfnamefont{O.}~\bibnamefont{Yarkoni}}, \bibnamefont{and}
		\bibinfo{author}{\bibfnamefont{D.}~\bibnamefont{Frankel}},
		\bibinfo{journal}{Soft Matter} \textbf{\bibinfo{volume}{11}},
		\bibinfo{pages}{7722} (\bibinfo{year}{2015}).
		
		\bibitem[{\citenamefont{Caspar and Klug}(1962)}]{Cas62}
		\bibinfo{author}{\bibfnamefont{D.~L.} \bibnamefont{Caspar}} \bibnamefont{and}
		\bibinfo{author}{\bibfnamefont{A.}~\bibnamefont{Klug}},
		\bibinfo{journal}{Cold Spring Harbor symposia on quantitative biology}
		\textbf{\bibinfo{volume}{27}}, \bibinfo{pages}{1—24}
		(\bibinfo{year}{1962}).
		
		\bibitem[{\citenamefont{\v{S}iber and Podgornik}(2009)}]{Sil09}
		\bibinfo{author}{\bibfnamefont{A.}~\bibnamefont{\v{S}iber}} \bibnamefont{and}
		\bibinfo{author}{\bibfnamefont{R.}~\bibnamefont{Podgornik}},
		\bibinfo{journal}{Phys. Rev. E} \textbf{\bibinfo{volume}{79}},
		\bibinfo{pages}{011919} (\bibinfo{year}{2009}).
		
		\bibitem[{\citenamefont{Ceres and Zlotnick}(2002)}]{Cer02}
		\bibinfo{author}{\bibfnamefont{P.}~\bibnamefont{Ceres}} \bibnamefont{and}
		\bibinfo{author}{\bibfnamefont{A.}~\bibnamefont{Zlotnick}},
		\bibinfo{journal}{Biochemistry} \textbf{\bibinfo{volume}{41}},
		\bibinfo{pages}{11525} (\bibinfo{year}{2002}).
		
		\bibitem[{\citenamefont{\v{S}iber and Podgornik}(2008)}]{Sil08}
		\bibinfo{author}{\bibfnamefont{A.}~\bibnamefont{\v{S}iber}} \bibnamefont{and}
		\bibinfo{author}{\bibfnamefont{R.}~\bibnamefont{Podgornik}},
		\bibinfo{journal}{Physical review. E, Statistical, nonlinear, and soft matter
			physics} \textbf{\bibinfo{volume}{76}}, \bibinfo{pages}{061906}
		(\bibinfo{year}{2008}).
		
		\bibitem[{\citenamefont{Loo et~al.}(2006)\citenamefont{Loo, Guenther,
				Basnayake, Lommel, and Franzen}}]{Loo06}
		\bibinfo{author}{\bibfnamefont{L.}~\bibnamefont{Loo}},
		\bibinfo{author}{\bibfnamefont{R.~H.} \bibnamefont{Guenther}},
		\bibinfo{author}{\bibfnamefont{V.~R.} \bibnamefont{Basnayake}},
		\bibinfo{author}{\bibfnamefont{S.~A.} \bibnamefont{Lommel}},
		\bibnamefont{and} \bibinfo{author}{\bibfnamefont{S.}~\bibnamefont{Franzen}},
		\bibinfo{journal}{Journal of the American Chemical Society}
		\textbf{\bibinfo{volume}{128}}, \bibinfo{pages}{4502} (\bibinfo{year}{2006}).
		
		\bibitem[{\citenamefont{Perlmutter
				et~al.}(2014{\natexlab{a}})\citenamefont{Perlmutter, Perkett, and
				Hagan}}]{Per14}
		\bibinfo{author}{\bibfnamefont{J.~D.} \bibnamefont{Perlmutter}},
		\bibinfo{author}{\bibfnamefont{M.~R.} \bibnamefont{Perkett}},
		\bibnamefont{and} \bibinfo{author}{\bibfnamefont{M.~F.} \bibnamefont{Hagan}},
		\bibinfo{journal}{Journal of Molecular Biology}
		\textbf{\bibinfo{volume}{426}}, \bibinfo{pages}{3148 }
		(\bibinfo{year}{2014}{\natexlab{a}}).
		
		\bibitem[{\citenamefont{Perlmutter
				et~al.}(2014{\natexlab{b}})\citenamefont{Perlmutter, Perkett, and
				Hagan}}]{Jas14}
		\bibinfo{author}{\bibfnamefont{J.~D.} \bibnamefont{Perlmutter}},
		\bibinfo{author}{\bibfnamefont{M.~R.} \bibnamefont{Perkett}},
		\bibnamefont{and} \bibinfo{author}{\bibfnamefont{M.~F.} \bibnamefont{Hagan}},
		\bibinfo{journal}{Journal of Molecular Biology}
		\textbf{\bibinfo{volume}{426}}, \bibinfo{pages}{3148 }
		(\bibinfo{year}{2014}{\natexlab{b}}).
		
		\bibitem[{\citenamefont{Hagan}(2009)}]{Hag09}
		\bibinfo{author}{\bibfnamefont{M.~F.} \bibnamefont{Hagan}},
		\bibinfo{journal}{The Journal of Chemical Physics}
		\textbf{\bibinfo{volume}{130}}, \bibinfo{pages}{114902}
		(\bibinfo{year}{2009}).
		
		\bibitem[{\citenamefont{Javidpour et~al.}(2013)\citenamefont{Javidpour,
				Bo\v{z}i\v{c}, Naji, and Podgornik}}]{Jav13}
		\bibinfo{author}{\bibfnamefont{L.}~\bibnamefont{Javidpour}},
		\bibinfo{author}{\bibfnamefont{A.~L.} \bibnamefont{Bo\v{z}i\v{c}}},
		\bibinfo{author}{\bibfnamefont{A.}~\bibnamefont{Naji}}, \bibnamefont{and}
		\bibinfo{author}{\bibfnamefont{R.}~\bibnamefont{Podgornik}},
		\bibinfo{journal}{Soft Matter} \textbf{\bibinfo{volume}{9}},
		\bibinfo{pages}{11357} (\bibinfo{year}{2013}).
		
		\bibitem[{\citenamefont{Messina}(2009)}]{Mes09}
		\bibinfo{author}{\bibfnamefont{R.}~\bibnamefont{Messina}},
		\bibinfo{journal}{Journal of Physics: Condensed Matter}
		\textbf{\bibinfo{volume}{21}}, \bibinfo{pages}{113102}
		(\bibinfo{year}{2009}).
		
		\bibitem[{\citenamefont{French et~al.}(2010)\citenamefont{French, Parsegian,
				Podgornik, Rajter, Jagota, Luo, Asthagiri, Chaudhury, Chiang, Granick
				et~al.}}]{Fre10}
		\bibinfo{author}{\bibfnamefont{R.~H.} \bibnamefont{French}},
		\bibinfo{author}{\bibfnamefont{V.~A.} \bibnamefont{Parsegian}},
		\bibinfo{author}{\bibfnamefont{R.}~\bibnamefont{Podgornik}},
		\bibinfo{author}{\bibfnamefont{R.~F.} \bibnamefont{Rajter}},
		\bibinfo{author}{\bibfnamefont{A.}~\bibnamefont{Jagota}},
		\bibinfo{author}{\bibfnamefont{J.}~\bibnamefont{Luo}},
		\bibinfo{author}{\bibfnamefont{D.}~\bibnamefont{Asthagiri}},
		\bibinfo{author}{\bibfnamefont{M.~K.} \bibnamefont{Chaudhury}},
		\bibinfo{author}{\bibfnamefont{Y.-m.} \bibnamefont{Chiang}},
		\bibinfo{author}{\bibfnamefont{S.}~\bibnamefont{Granick}},
		\bibnamefont{et~al.}, \bibinfo{journal}{Rev. Mod. Phys.}
		\textbf{\bibinfo{volume}{82}}, \bibinfo{pages}{1887} (\bibinfo{year}{2010}).
		
		\bibitem[{\citenamefont{Colla et~al.}(2016)\citenamefont{Colla, Girotto, dos
				Santos, and Levin}}]{Col16}
		\bibinfo{author}{\bibfnamefont{T.}~\bibnamefont{Colla}},
		\bibinfo{author}{\bibfnamefont{M.}~\bibnamefont{Girotto}},
		\bibinfo{author}{\bibfnamefont{A.~P.} \bibnamefont{dos Santos}},
		\bibnamefont{and} \bibinfo{author}{\bibfnamefont{Y.}~\bibnamefont{Levin}},
		\bibinfo{journal}{The Journal of Chemical Physics}
		\textbf{\bibinfo{volume}{145}}, \bibinfo{pages}{094704}
		(\bibinfo{year}{2016}).
		
		\bibitem[{\citenamefont{Zandi et~al.}(2006)\citenamefont{Zandi, van~der Schoot,
				Reguera, Kegel, and Reiss}}]{Zan06}
		\bibinfo{author}{\bibfnamefont{R.}~\bibnamefont{Zandi}},
		\bibinfo{author}{\bibfnamefont{P.}~\bibnamefont{van~der Schoot}},
		\bibinfo{author}{\bibfnamefont{D.}~\bibnamefont{Reguera}},
		\bibinfo{author}{\bibfnamefont{W.}~\bibnamefont{Kegel}}, \bibnamefont{and}
		\bibinfo{author}{\bibfnamefont{H.}~\bibnamefont{Reiss}},
		\bibinfo{journal}{Biophysical journal} \textbf{\bibinfo{volume}{90}},
		\bibinfo{pages}{1939} (\bibinfo{year}{2006}).
		
		\bibitem[{\citenamefont{Ninham and Parsegian}(1971)}]{Nin71}
		\bibinfo{author}{\bibfnamefont{B.~W.} \bibnamefont{Ninham}} \bibnamefont{and}
		\bibinfo{author}{\bibfnamefont{V.}~\bibnamefont{Parsegian}},
		\bibinfo{journal}{Journal of Theoretical Biology}
		\textbf{\bibinfo{volume}{31}}, \bibinfo{pages}{405 } (\bibinfo{year}{1971}).
		
		\bibitem[{\citenamefont{Trefalt et~al.}(2016)\citenamefont{Trefalt, Behrens,
				and Borkovec}}]{Tre16}
		\bibinfo{author}{\bibfnamefont{G.}~\bibnamefont{Trefalt}},
		\bibinfo{author}{\bibfnamefont{S.~H.} \bibnamefont{Behrens}},
		\bibnamefont{and} \bibinfo{author}{\bibfnamefont{M.}~\bibnamefont{Borkovec}},
		\bibinfo{journal}{Langmuir} \textbf{\bibinfo{volume}{32}},
		\bibinfo{pages}{380} (\bibinfo{year}{2016}).
		
		\bibitem[{\citenamefont{Markovich et~al.}(2017)\citenamefont{Markovich,
				Andelman, and Podgornik}}]{Mar17}
		\bibinfo{author}{\bibfnamefont{T.}~\bibnamefont{Markovich}},
		\bibinfo{author}{\bibfnamefont{D.}~\bibnamefont{Andelman}}, \bibnamefont{and}
		\bibinfo{author}{\bibfnamefont{R.}~\bibnamefont{Podgornik}},
		\bibinfo{journal}{{EPL} (Europhysics Letters)}
		\textbf{\bibinfo{volume}{120}}, \bibinfo{pages}{26001}
		(\bibinfo{year}{2017}).
		
		\bibitem[{\citenamefont{Podgornik}(2018)}]{Pod18}
		\bibinfo{author}{\bibfnamefont{R.}~\bibnamefont{Podgornik}},
		\bibinfo{journal}{The Journal of Chemical Physics}
		\textbf{\bibinfo{volume}{149}}, \bibinfo{pages}{104701}
		(\bibinfo{year}{2018}).
		
		\bibitem[{\citenamefont{Smith et~al.}(2018)\citenamefont{Smith, Maroni,
				Borkovec, and Trefalt}}]{Smi18}
		\bibinfo{author}{\bibfnamefont{A.~M.} \bibnamefont{Smith}},
		\bibinfo{author}{\bibfnamefont{P.}~\bibnamefont{Maroni}},
		\bibinfo{author}{\bibfnamefont{M.}~\bibnamefont{Borkovec}}, \bibnamefont{and}
		\bibinfo{author}{\bibfnamefont{G.}~\bibnamefont{Trefalt}},
		\bibinfo{journal}{Colloids Interfaces} \textbf{\bibinfo{volume}{2}},
		\bibinfo{pages}{65} (\bibinfo{year}{2018}).
		
		\bibitem[{\citenamefont{Frydel}(2019)}]{Fry19}
		\bibinfo{author}{\bibfnamefont{D.}~\bibnamefont{Frydel}}, \bibinfo{journal}{The
			Journal of Chemical Physics} \textbf{\bibinfo{volume}{150}},
		\bibinfo{pages}{194901} (\bibinfo{year}{2019}).
		
		\bibitem[{\citenamefont{Bakhshandeh et~al.}(2019)\citenamefont{Bakhshandeh,
				Frydel, Diehl, and Levin}}]{Bak19}
		\bibinfo{author}{\bibfnamefont{A.}~\bibnamefont{Bakhshandeh}},
		\bibinfo{author}{\bibfnamefont{D.}~\bibnamefont{Frydel}},
		\bibinfo{author}{\bibfnamefont{A.}~\bibnamefont{Diehl}}, \bibnamefont{and}
		\bibinfo{author}{\bibfnamefont{Y.}~\bibnamefont{Levin}},
		\bibinfo{journal}{Phys. Rev. Lett.} \textbf{\bibinfo{volume}{123}},
		\bibinfo{pages}{208004} (\bibinfo{year}{2019}).
		
		\bibitem[{\citenamefont{\v{S}amaj and Trizac}(2011)}]{Sam11}
		\bibinfo{author}{\bibfnamefont{L.}~\bibnamefont{\v{S}amaj}} \bibnamefont{and}
		\bibinfo{author}{\bibfnamefont{E.}~\bibnamefont{Trizac}},
		\bibinfo{journal}{Phys. Rev. Lett.} \textbf{\bibinfo{volume}{106}},
		\bibinfo{pages}{078301} (\bibinfo{year}{2011}).
		
		\bibitem[{\citenamefont{\v{S}amaj et~al.}(2016)\citenamefont{\v{S}amaj, dos
				Santos, Levin, and Trizac}}]{Sam16_2}
		\bibinfo{author}{\bibfnamefont{L.}~\bibnamefont{\v{S}amaj}},
		\bibinfo{author}{\bibfnamefont{A.~P.} \bibnamefont{dos Santos}},
		\bibinfo{author}{\bibfnamefont{Y.}~\bibnamefont{Levin}}, \bibnamefont{and}
		\bibinfo{author}{\bibfnamefont{E.}~\bibnamefont{Trizac}},
		\bibinfo{journal}{Soft Matter} \textbf{\bibinfo{volume}{12}},
		\bibinfo{pages}{8768} (\bibinfo{year}{2016}).
		
		\bibitem[{\citenamefont{\v{S}amaj and Trizac}(2016)}]{Sam16}
		\bibinfo{author}{\bibfnamefont{L.}~\bibnamefont{\v{S}amaj}} \bibnamefont{and}
		\bibinfo{author}{\bibfnamefont{E.}~\bibnamefont{Trizac}},
		\bibinfo{journal}{Phys. Rev. E} \textbf{\bibinfo{volume}{93}},
		\bibinfo{pages}{012601} (\bibinfo{year}{2016}).
		
		\bibitem[{\citenamefont{Hansen and McDonald}(2006)}]{Hansen}
		\bibinfo{author}{\bibfnamefont{J.~P.} \bibnamefont{Hansen}} \bibnamefont{and}
		\bibinfo{author}{\bibfnamefont{I.~R.} \bibnamefont{McDonald}},
		\emph{\bibinfo{title}{Theory of Simple Liquids}}
		(\bibinfo{publisher}{Academic Press}, \bibinfo{address}{London},
		\bibinfo{year}{2006}).
		
		\bibitem[{\citenamefont{Yang and Liu}(2015)}]{Yan15}
		\bibinfo{author}{\bibfnamefont{G.}~\bibnamefont{Yang}} \bibnamefont{and}
		\bibinfo{author}{\bibfnamefont{L.}~\bibnamefont{Liu}}, \bibinfo{journal}{The
			Journal of Chemical Physics} \textbf{\bibinfo{volume}{142}},
		\bibinfo{pages}{194110} (\bibinfo{year}{2015}).
		
		\bibitem[{\citenamefont{Rosenfeld}(1993)}]{Ros93}
		\bibinfo{author}{\bibfnamefont{Y.}~\bibnamefont{Rosenfeld}},
		\bibinfo{journal}{The Journal of Chemical Physics}
		\textbf{\bibinfo{volume}{98}}, \bibinfo{pages}{8126} (\bibinfo{year}{1993}).
		
		\bibitem[{\citenamefont{Jiang et~al.}(2014)\citenamefont{Jiang, Cao, Henderson,
				and Wu}}]{Jia14}
		\bibinfo{author}{\bibfnamefont{J.}~\bibnamefont{Jiang}},
		\bibinfo{author}{\bibfnamefont{D.}~\bibnamefont{Cao}},
		\bibinfo{author}{\bibfnamefont{D.}~\bibnamefont{Henderson}},
		\bibnamefont{and} \bibinfo{author}{\bibfnamefont{J.}~\bibnamefont{Wu}},
		\bibinfo{journal}{The Journal of Chemical Physics}
		\textbf{\bibinfo{volume}{140}}, \bibinfo{pages}{044714}
		(\bibinfo{year}{2014}).
		
		\bibitem[{\citenamefont{Rossmann and Johnson}(1989)}]{Ros89}
		\bibinfo{author}{\bibfnamefont{M.~G.} \bibnamefont{Rossmann}} \bibnamefont{and}
		\bibinfo{author}{\bibfnamefont{J.~E.} \bibnamefont{Johnson}},
		\bibinfo{journal}{Annual Review of Biochemistry}
		\textbf{\bibinfo{volume}{58}}, \bibinfo{pages}{533} (\bibinfo{year}{1989}).
		
		\bibitem[{\citenamefont{Rosenfeld}(1990)}]{Ros90}
		\bibinfo{author}{\bibfnamefont{Y.}~\bibnamefont{Rosenfeld}},
		\bibinfo{journal}{The Journal of Chemical Physics}
		\textbf{\bibinfo{volume}{93}}, \bibinfo{pages}{4305} (\bibinfo{year}{1990}).
		
		\bibitem[{\citenamefont{Roth}(2010)}]{Rot10}
		\bibinfo{author}{\bibfnamefont{R.}~\bibnamefont{Roth}},
		\bibinfo{journal}{Journal of Physics: Condensed Matter}
		\textbf{\bibinfo{volume}{22}}, \bibinfo{pages}{063102}
		(\bibinfo{year}{2010}).
		
		\bibitem[{\citenamefont{Roth et~al.}(2002)\citenamefont{Roth, Evans, Lang, and
				Kahl}}]{Rot02}
		\bibinfo{author}{\bibfnamefont{R.}~\bibnamefont{Roth}},
		\bibinfo{author}{\bibfnamefont{R.}~\bibnamefont{Evans}},
		\bibinfo{author}{\bibfnamefont{A.}~\bibnamefont{Lang}}, \bibnamefont{and}
		\bibinfo{author}{\bibfnamefont{G.}~\bibnamefont{Kahl}},
		\bibinfo{journal}{Journal of Physics: Condensed Matter}
		\textbf{\bibinfo{volume}{14}}, \bibinfo{pages}{12063} (\bibinfo{year}{2002}).
		
		\bibitem[{\citenamefont{Davidchack et~al.}(2016)\citenamefont{Davidchack,
				Laird, and Roth}}]{Dav16}
		\bibinfo{author}{\bibfnamefont{R.~L.} \bibnamefont{Davidchack}},
		\bibinfo{author}{\bibfnamefont{B.~B.} \bibnamefont{Laird}}, \bibnamefont{and}
		\bibinfo{author}{\bibfnamefont{R.}~\bibnamefont{Roth}},
		\bibinfo{journal}{Condensed Matter Physics} \textbf{\bibinfo{volume}{19}},
		\bibinfo{pages}{23001} (\bibinfo{year}{2016}).
		
		\bibitem[{\citenamefont{Zhou and Jamnik}(2005)}]{Zho05}
		\bibinfo{author}{\bibfnamefont{S.}~\bibnamefont{Zhou}} \bibnamefont{and}
		\bibinfo{author}{\bibfnamefont{A.}~\bibnamefont{Jamnik}},
		\bibinfo{journal}{The Journal of Chemical Physics}
		\textbf{\bibinfo{volume}{122}}, \bibinfo{pages}{064503}
		(\bibinfo{year}{2005}).
		
		\bibitem[{\citenamefont{Colla et~al.}(2017)\citenamefont{Colla, Nunes~Lopes,
				and dos Santos}}]{Col17}
		\bibinfo{author}{\bibfnamefont{T.}~\bibnamefont{Colla}},
		\bibinfo{author}{\bibfnamefont{L.}~\bibnamefont{Nunes~Lopes}},
		\bibnamefont{and} \bibinfo{author}{\bibfnamefont{A.~P.} \bibnamefont{dos
				Santos}}, \bibinfo{journal}{The Journal of Chemical Physics}
		\textbf{\bibinfo{volume}{147}}, \bibinfo{pages}{014104}
		(\bibinfo{year}{2017}).
		
		\bibitem[{\citenamefont{Allen and Tildesley}(2017)}]{Allen}
		\bibinfo{author}{\bibfnamefont{M.~P.} \bibnamefont{Allen}} \bibnamefont{and}
		\bibinfo{author}{\bibfnamefont{D.~J.} \bibnamefont{Tildesley}},
		\emph{\bibinfo{title}{Computer Simulation of Liquids, 2nd}}
		(\bibinfo{publisher}{Oxford University Press, Inc. New York},
		\bibinfo{year}{2017}).
		
		\bibitem[{\citenamefont{Bakhshandeh}(2018)}]{bkh18}
		\bibinfo{author}{\bibfnamefont{A.}~\bibnamefont{Bakhshandeh}},
		\bibinfo{journal}{Chemical Physics} \textbf{\bibinfo{volume}{513}},
		\bibinfo{pages}{195} (\bibinfo{year}{2018}).
		
		\bibitem[{\citenamefont{Colla et~al.}(2014)\citenamefont{Colla, Likos, and
				Levin}}]{Col14}
		\bibinfo{author}{\bibfnamefont{T.}~\bibnamefont{Colla}},
		\bibinfo{author}{\bibfnamefont{C.~N.} \bibnamefont{Likos}}, \bibnamefont{and}
		\bibinfo{author}{\bibfnamefont{Y.}~\bibnamefont{Levin}},
		\bibinfo{journal}{The Journal of Chemical Physics}
		\textbf{\bibinfo{volume}{141}}, \bibinfo{pages}{234902}
		(\bibinfo{year}{2014}).
		
		\bibitem[{\citenamefont{Denton and Tang}(2016)}]{Den16}
		\bibinfo{author}{\bibfnamefont{A.~R.} \bibnamefont{Denton}} \bibnamefont{and}
		\bibinfo{author}{\bibfnamefont{Q.}~\bibnamefont{Tang}}, \bibinfo{journal}{The
			Journal of Chemical Physics} \textbf{\bibinfo{volume}{145}},
		\bibinfo{pages}{164901} (\bibinfo{year}{2016}).
		
		\bibitem[{\citenamefont{Tergolina and dos Santos}(2017)}]{Ter17}
		\bibinfo{author}{\bibfnamefont{V.~B.} \bibnamefont{Tergolina}}
		\bibnamefont{and} \bibinfo{author}{\bibfnamefont{A.~P.} \bibnamefont{dos
				Santos}}, \bibinfo{journal}{The Journal of Chemical Physics}
		\textbf{\bibinfo{volume}{147}}, \bibinfo{pages}{114103}
		(\bibinfo{year}{2017}).
		
		\bibitem[{\citenamefont{Evans and Needham}(1987)}]{Eva87}
		\bibinfo{author}{\bibfnamefont{E.}~\bibnamefont{Evans}} \bibnamefont{and}
		\bibinfo{author}{\bibfnamefont{D.}~\bibnamefont{Needham}},
		\bibinfo{journal}{The Journal of Physical Chemistry}
		\textbf{\bibinfo{volume}{91}}, \bibinfo{pages}{4219} (\bibinfo{year}{1987}).
		
		\bibitem[{\citenamefont{Landau et~al.}(2012)\citenamefont{Landau, Pitaevskii,
				Kosevich, and Lifshitz}}]{Landau}
		\bibinfo{author}{\bibfnamefont{L.~D.} \bibnamefont{Landau}},
		\bibinfo{author}{\bibfnamefont{L.~P.} \bibnamefont{Pitaevskii}},
		\bibinfo{author}{\bibfnamefont{A.~M.} \bibnamefont{Kosevich}},
		\bibnamefont{and} \bibinfo{author}{\bibfnamefont{E.}~\bibnamefont{Lifshitz}},
		\emph{\bibinfo{title}{Theory of Elasticity}} (\bibinfo{publisher}{London:
			Butterworth-Heinemann}, \bibinfo{year}{2012}).
		
		\bibitem[{\citenamefont{Mallarino et~al.}(2015)\citenamefont{Mallarino,
				T\'ellez, and Trizac}}]{Mal15}
		\bibinfo{author}{\bibfnamefont{J.~P.} \bibnamefont{Mallarino}},
		\bibinfo{author}{\bibfnamefont{G.}~\bibnamefont{T\'ellez}}, \bibnamefont{and}
		\bibinfo{author}{\bibfnamefont{E.}~\bibnamefont{Trizac}},
		\bibinfo{journal}{Molecular Physics} \textbf{\bibinfo{volume}{113}},
		\bibinfo{pages}{2409} (\bibinfo{year}{2015}).
		
		\bibitem[{\citenamefont{Reddy et~al.}(1998)\citenamefont{Reddy, Giesing,
				Morton, Kumar, Post, Brooks, and Johnson}}]{Vij98}
		\bibinfo{author}{\bibfnamefont{V.~S.} \bibnamefont{Reddy}},
		\bibinfo{author}{\bibfnamefont{H.~A.} \bibnamefont{Giesing}},
		\bibinfo{author}{\bibfnamefont{R.~T.} \bibnamefont{Morton}},
		\bibinfo{author}{\bibfnamefont{A.}~\bibnamefont{Kumar}},
		\bibinfo{author}{\bibfnamefont{C.~B.} \bibnamefont{Post}},
		\bibinfo{author}{\bibfnamefont{C.~L.} \bibnamefont{Brooks}},
		\bibnamefont{and} \bibinfo{author}{\bibfnamefont{J.~E.}
			\bibnamefont{Johnson}}, \bibinfo{journal}{Biophysical Journal}
		\textbf{\bibinfo{volume}{74}}, \bibinfo{pages}{546 } (\bibinfo{year}{1998}),
		ISSN \bibinfo{issn}{0006-3495}.
		
		\bibitem[{\citenamefont{Trizac and Shen}(2016)}]{Tri16}
		\bibinfo{author}{\bibfnamefont{E.}~\bibnamefont{Trizac}} \bibnamefont{and}
		\bibinfo{author}{\bibfnamefont{T.}~\bibnamefont{Shen}},
		\bibinfo{journal}{{EPL} (Europhysics Letters)}
		\textbf{\bibinfo{volume}{116}}, \bibinfo{pages}{18007}
		(\bibinfo{year}{2016}).
		
	\end{thebibliography}
\end{document}